\documentclass{aastex61}

\newcommand\aastex{AAS\TeX}

\shorttitle{\aastex\ sample article}
\shortauthors{Gribel, Miranda \& Vilas-Boas}

\begin{document}				

\title{Connecting the Cosmic Star Formation Rate with the Local Star Formation}

\correspondingauthor{Carolina Gribel}
\email{carol.gribel@gmail.com, oswaldo.miranda@inpe.br, williams.boas@inpe.br}

\author{Carolina Gribel}, \author{Oswaldo D. Miranda}, \author{Jos{\' e} Williams Vilas-Boas}
\affiliation{Instituto Nacional de Pesquisas Espaciais---INPE, Divis{\~ a}o de Astrof{\'i}sica, Av. dos Astronautas, 1.758---Jardim da Granja, S{\~ a}o Jos{\' e} dos Campos---SP, Brazil}

\begin{abstract}

We present a model that unifies the cosmic star formation rate (CSFR), obtained through the hierarchical structure formation scenario, with the (Galactic) local star formation rate (SFR). It is possible to use the SFR to generate a CSFR mapping through the density probability distribution functions (PDFs) commonly used to study the role of turbulence in the star-forming regions of the Galaxy. We obtain a consistent mapping from redshift $z\sim 20$ up to the present ($z = 0$). Our results show that the turbulence exhibits a dual character, providing high values for the star formation efficiency ($\langle\varepsilon\rangle \sim 0.32$) in the redshift interval $z\sim 3.5-20$ and reducing its value to $\langle\varepsilon\rangle = 0.021$ at $z = 0$. The value of the Mach number ($\mathcal{M}_{\rm crit}$), from which $\langle\varepsilon\rangle$ rapidly decreases, is dependent on both the polytropic index ($\Gamma$) and the minimum density contrast of the gas. We also derive Larson's first law associated with the velocity dispersion ($\langle V_{\rm rms}\rangle$) in the local star formation regions. Our model shows good agreement with Larson's law in the $\sim 10-50\,{\rm pc}$ range, providing typical temperatures $T_{0} \sim 10-80\,{\rm K}$ for the gas associated with star formation. As a consequence, dark matter halos of great mass could contain a number of halos of much smaller mass, and be able to form structures similar to globular clusters. Thus, Larson's law emerges as a result of the very formation of large-scale structures, which in turn would allow the formation of galactic systems, including our Galaxy.

\end{abstract}


\section{Introduction} \label{sec:intro}

Understanding how galaxies form in the universe is certainly one of the main goals of modern cosmology. The formation of galaxies is a process intrinsically related to the evolution of cosmological structures (for a recent review on this subject see, e.g., \citeauthor{frenkwhite12} \citeyear{frenkwhite12}). In particular, after the radiation-baryonic matter decoupling, which occurred in redshift $z \sim 1100$, the density perturbations, generated during the inflationary phase of the universe, are able to grow more enhanced by the action of their self-gravity producing the so-called halos of dark matter. These dark matter structures generate potential wells that allow them to capture the baryonic matter of the surrounding environment, initiating the production of stars at some time near redshift 20. This is, roughly speaking, the process that leads to the formation of large-scale structures of the universe. Within the cosmological context, both theoretically and observationally, star formation is described by the so-called cosmic star formation rate (CSFR), represented as a function of redshift in units of $M_{\odot}\,{\rm Mpc^{-3}\,yr^{-1}}$. The current status of CSFR in both theoretical and observational aspects can be found in a recent article by \cite{madau_csfr}.

On the other hand, our knowledge about the processes associated with star formation at the local (galactic) level begins with the work of \cite{schmidt1959,schmidt1963}, whose objective was to find a correlation between gas surface density in galaxies and the stellar formation rate. \cite{Kennicutt1998} used Schmidt's power-law function to fit observational data of disk and starburst galaxies and to determine the best-fit slope and normalization. The derived relation can be represented as

\begin{equation}
\Sigma_{\rm SFR} = (2.5\pm 0.7)\times 10^{-4}\left(\frac{\Sigma_{\rm gas}}{1M_{\odot}\,{\rm pc}^{-2}}\right)^{1.4\pm 0.15}M_{\odot}\,{\rm kpc^{-2}\,yr^{-1}},
\label{sfr_local}
\end{equation}

\noindent where $\Sigma_{\rm SFR}$ is the star formation rate (SFR) per unit area and $\Sigma_{\rm gas}$ is the gas surface density. This correlation can be applied to a large number of nearby galaxies. Although \cite{Kennicutt1998} found that the relationship was adjusted by the exponent ${1.4\pm 0.15}$, a similar result could be obtained by dividing $\Sigma_{\rm gas}$, in Equation (\ref{sfr_local}), by $\tau_{\rm dyn}$---the disk orbital time (see, \citeauthor{Kennicutt1998} \citeyear{Kennicutt1998}; \citeauthor{martin2001} \citeyear{martin2001}).

However, as highlighted by \cite{salim2015}, a significant scattering remains from these scenarios, so that $\Sigma_{\rm SFR}$ can vary significantly for any of the two inputs, i.e., $\Sigma_{\rm gas}$ and $\Sigma_{\rm gas}/\tau$ (see also \citeauthor{heiderman2010} \citeyear{heiderman2010}; \citeauthor{krumholz2012} \citeyear{krumholz2012}; \citeauthor{federrath2013a} \citeyear{federrath2013a}). Additionally, with the improvement of observational data over the last 20 years, especially through CO observations, it has been possible to study the correlation between molecular gas and SFR at scales $\sim 0.5-1\,{\rm kpc}$ \citep{leroy2013}. In particular, this correlation has shown that the depletion time is approximately constant with $\tau_{\rm dep} = \Sigma_{\rm mol}/\Sigma_{\rm SFR} \approx 2.2\,{\rm Gyr}$, where $\tau_{\rm dep}$ is the time required for the star formation to use up the current molecular gas supply. It is important to note that there is some controversy in the literature about the constancy of $\tau_{\rm dep}$. For example, using COLD GASS data (CO LEGACY DATABASE FOR GASS---Galex Arecibo SDSS Survey), \citeauthor{saintonge2011a} (\citeyear{saintonge2011a}, \citeyear{saintonge2011b}) find a non-constant depletion time over a wide range of galaxies, although the variation is small.

Based on the mass of molecular gas within the solar circle, which is on the order of $10^{9}M_{\sun}$, and the SFR in the Galaxy ($\sim 1\,M_{\sun}{\rm yr}^{-1}$, so yielding $\tau_{\rm dep} \sim 1\,{\rm Gyr}$, corresponding to 100 times the freefall time), \cite{krumholz2005} suggest that the ratio $\varepsilon_{\rm ff} = \tau_{\rm ff} / \tau_{\rm dep} \sim 1/100$ could provide an observational constraint to stellar formation theories. In particular, $\varepsilon_{\rm ff}$ is called the  dimensionless star formation rate per freefall time. This quantity represents the mass of molecular gas converted into stars per freefall time of the system.

The low inferred value for $\varepsilon_{\rm ff}$ surely raises the question of what is the main factor that makes the star formation rate so small in molecular clouds. Although it is possible to consider different mechanisms to explain this result, turbulence has the greatest potential to regulate star formation. In particular, interstellar turbulence as a key for star formation has been studied for a long time (see, e.g., \citeauthor{klessen2000} \citeyear{klessen2000}; \citeauthor{elmegreen2004} \citeyear{elmegreen2004}; \citeauthor{krumholz2005} \citeyear{krumholz2005}; \citeauthor{mckee2007} \citeyear{mckee2007}) and has been successively refined and improved by several authors in the last years (see, e.g., \citeauthor{hennebelle2011} \citeyear{hennebelle2011}; \citeauthor{padoan2011} \citeyear{padoan2011}; \citeauthor{federrath2012} \citeyear{federrath2012}; \citeauthor{krumholz2012} \citeyear{krumholz2012}; \citeauthor{kritsuk2013} \citeyear{kritsuk2013}; \citeauthor{padoan2014} \citeyear{padoan2014}). It is also important to highlight the recent study by \cite{federrath2015a} showing that only the combination of turbulence, magnetic fields, and protostellar feedback (through jets and outflows) yields realistic (low) SFRs in that observed range of a few percent per freefall time.

Additionally, it has been suggested that turbulence could play a dual role in star formation. In particular, \cite{klessen2010} and \cite{klessen2011} remark that this duality would come from the fact that turbulence provides support on a global scale, but can promote collapse on a local scale. As a consequence, the birth of a star is dynamically connected with the parental gas cloud, thus determining when and where a protostar forms.

\cite{klessen2010} have also pointed out that the role of turbulence in the formation of the first stars in the universe, which put an end to the so-called ``dark ages," is less understood. In general, the formation of the first stars is studied through numerical simulations involving the collapse and virialization ($T_{\rm vir}\sim 10^{3}{\rm K}$) of dark matter halos at redshift $z\sim  20 - 30$, which generated the conditions for star formation (see, e.g., \citeauthor{bromm2004} \citeyear{bromm2004}; \citeauthor{bromm2009} \citeyear{bromm2009}). Notwithstanding, the results of \cite{klessen2010} indicated that the first stars of the universe were subject to the same dynamic processes of the local star-forming regions. In addition, the simulations carried out by the authors have shown that the mass function of the primordial protostars should be comparable to the present-day initial mass function (IMF) (see also the recent results on primordial protostars in \citeauthor{dutta2015} \citeyear{dutta2015} and \citeauthor{hosokawa2016} \citeyear{hosokawa2016}).

The objective of the present work is to show the complementarity between the formulation used to obtain the CSFR and the modeling used to characterize the local rate of star formation (described by the SFR). Furthermore, the SFR can provide the CSFR with an estimate of the turbulence, through the Mach number, as a function of redshift. On the other hand, the CSFR can provide the SFR with a way of naturally obtaining the Larson's first law that associates the internal motions with the structure of the molecular clouds where the star formation takes place. In principle, our model may give some clues about the dual role of turbulence in star formation as initially suggested by \cite{klessen2010}.

This paper is organized as follows: in section \ref{sec:sfr}, we review the model of \cite{pereira2010}, hereinafter referred to as PM, that allows deriving the CSFR that will be the cosmological basis of our work. Also in section \ref{sec:sfr}, we review the main points discussed in the literature on the SFR in order to better characterize our {\it ansatz} on the complementarity between CSFR and SFR. In particular, the characterization of the SFR, to be compared with the CSFR, will be based on the works of \cite{hopkins2013a} and \cite{federrath2015}, hereinafter referred to as H13 and FB15, respectively. In section \ref{results}, we present our main results, and we present a summary and our conclusions in section \ref{sec:conc}.

\section{Scenarios for star formation} \label{sec:sfr}

\subsection{Cosmic Star Formation Rate---CSFR} \label{subsec:csfr}

PM used a Press--Schechter-like formalism to describe the formation of dark matter halos as a function of the redshift. The formation of these dark halos created the conditions for the baryonic matter of the cosmological environment to fall into the gravitational wells, seeding the birth of the first stars and thus contributing to the formation of large-scale structures of the universe. The authors coupled the star formation to this hierarchical (Press--Schechter) scenario through the laws of Schmidt and Salpeter. Thus, the CSFR can be obtained from redshift 20 to the present time, showing good agreement with the observational data within the range $0-5$ in redshift.

Our choice for PM-CSFR is based on its healthy applications. For example, \cite{pereira2011} analyzed different CSFRs discussed in the literature; their comparisons identify the PM-CSFR as the one that allows better adjustment with the inferred quasar luminosity function. Based on these results, the authors showed that the PM-CSFR could be directly connected with the growth of the supermassive black holes observed in the centers of most galaxies.

On the other hand, \cite{hao2013a} (see also \citeauthor{wei2016} \citeyear{wei2016}; \citeauthor{wei2017} \citeyear{wei2017}) showed that PM-CSFR can reproduce very well the cumulative function of Long Gamma Ray Bursts - LGRBs from redshift $z=0$ up to $z\sim 8$. These authors used a Kolmogorov--Smirnov test, which showed that PM-CSFR presents $p$-value $\sim 0.92$; that is, much better than the other CSFRs discussed in the literature. After that, \cite{hao2013b} used this CSFR to investigate the delay-time distribution of short GRB progenitors, which is an important property to constrain the progenitor of these sources (see also \citeauthor{wanderman2015} \citeyear{wanderman2015}).

The PM-CSFR is formulated on the scenario developed by \cite{ps74} who heuristically derived a mass function for bound virialized objects. The basic idea of this approach consists in defining halos as concentrations of mass that have already left the linear regime by crossing the threshold $\delta_{\rm c}$ for nonlinear collapse. Once the spectrum of fluctuations (power spectrum) is defined, it becomes relatively straightforward to calculate the halo mass function as a function of the mass and redshift. Thus, we can introduce the scale differential mass function $f(\sigma,z)$ (see \citeauthor{j1} \citeyear{j1}), defined as the fraction of the total mass per $\ln \sigma^{-1}$ that belongs to halos. That is,

\begin{equation}
f(\sigma,z)\equiv\frac{d\rho/\rho_{\rm B}}{d\ln\sigma^{-1}}=\frac{M}{\rho_{\rm B}(z)}\frac{dn(M,z)}{d\ln[\sigma^{-1}(M,z)]},\label{st}
\end{equation}

\noindent where $\rho$ is the dark matter halo density, $n(M,z)$ represents the number density of halos with mass $M$, $\rho_{\rm B}(z)$ is the background density (dark matter component) at redshift $z$, and $\sigma(M,z)$ is the variance of the linear density field. As highlighted in the work of \citet{j1}, this definition of the mass function has the advantage that it does not explicitly depend on redshift, power spectrum, or cosmology; all of these are contained in $\sigma(M,z)$ (see also \citealt{luk}). To determine $\sigma(M,z)$, the power spectrum $P(k)$ is smoothed with a spherical top-hat filter function of radius $R$, which on average encloses a mass $M$ $(R=[3M/4\pi\rho_{\rm B}(z)]^{1/3})$. In this way,

\begin{equation}
\sigma^{2}(M,z) = \frac{D^{2}(z)}{2\pi^{2}} \int_{0}^{\infty}{k^{2}P(k)W^{2}(k,M)dk},
\end{equation}

\noindent where $W(k,M)$ is the top-hat filter in the $k$-space

\begin{equation}
W(k,M) = \frac{3}{(kR)^{3}}[\sin(kR)-k R\cos(kR)],
\end{equation}

\noindent and the redshift dependence enters only through the growth factor $D(z)$. That is, $\sigma(M,z)=\sigma(M,0)D(z)$. The growth function can be approximated by \citep{carrol1992}:

\begin{equation}
D(a)\approx \frac{5 \Omega_{\rm m}(a)\ a}{2[1- \Omega_{\Lambda}(a)+\Omega_{\rm m}^{4/7}+\frac{1}{2}\Omega_{\rm m}(a)]},
\end{equation}

\noindent where the relative density of the ${\rm i}$-component is given by $\Omega_{\rm i}=\rho_{\rm i}/\rho_{\rm c}$, and ``${\rm i}$" representing dark energy ($\Lambda)$, and total matter (m), where total matter is the sum of baryonic matter (b) and dark matter (dm), while $a=1/(1+z)$ is the cosmological scale factor.

The primordial power spectrum has a power-law dependence on scale, that is, $P(k)\propto k^{n_{\rm p}}$. For a scale-invariant spectrum, the spectral index as predicted by inflation is $n_{\rm p}=1$. The current observational best fit for the spectral index is $n_{\rm p}=0.9667\pm 0.0044$ obtained from the data generated by the {\it Planck} satellite \citep{ade2014,ade2016}. The rate at which fluctuations grow on different scales is determined by an interplay between self-gravitation, pressure support, and damping processes. These effects lead to a modification of the form of the primordial power spectrum that is expressed in terms of a transfer function $T(k)$ given by:

\begin{equation}
P(k) = BkT(k),
\end{equation}
\noindent where the normalization factor $B$ is taken from observational data. For the transfer function, we consider \citep{e1}

\begin{equation}
T(k) = \frac{1}{\{1+[ak+(bk)^{3/2}+(ck)^{2}]^{\nu}\}^{2/\nu}},
\end{equation}

\noindent with $\nu = 1.13$, $a = (6.4 /\Gamma) h^{-1}\rm{Mpc}$, $b= (3.0/ \Gamma)h^{-1}\rm{Mpc}$, and $c = (1.7 /\Gamma) h^{-1}\rm{Mpc}$, where $\Gamma = \Omega_{\rm m} h\,\,{\rm e}^{-\Omega_{\rm b}(1+\sqrt{2h}/\Omega_{\rm m})}$ is the so-called shape parameter\footnote{In sections \ref{subsec:sfr} and \ref{results} we will use the same symbol $\Gamma$ to represent another physical quantity---the polytropic index.} of the power spectrum \citep{b2}. For the mass function presented in Equation (\ref{st}), we use the fit proposed by \cite{sheth1999}. That is,

\begin{equation}
f(\sigma,z) = 0.3222 \sqrt{\frac{2a}{\pi}} \frac{\delta_{\rm c}}{\sigma} \exp{\left(-\frac{a \delta_{c}^{2}}{2 \sigma^{2}} \right)} \left[1+\left(\frac{\sigma^{2}}{a\delta_{\rm c}^{2}}\right)^{p}\right]\label{est},
\end{equation}

\noindent where $\delta_{\rm c}=1.69$, while $a = 0.707$ and $p=0.3$. 

The parameterization of \cite{sheth1999} incorporates the possibility that the collapse of the halos is ellipsoidal--not only spherical, as proposed by \cite{ps74}. In addition, the \cite{sheth1999} fit has a very close agreement with numerical $N$-body simulations within a broad mass spectrum. With these considerations, we can determine the fraction of baryons that are incorporated into the halos as a function of both mass and redshift

\begin{equation}
f_{\rm b}(z)=\frac{\int_{M_{\rm min}}^{\infty} {f(\sigma)MdM}}{\int_{0}^{\infty} {f(\sigma)MdM}}.\label{fbaryon}
\end{equation}

The fact that  stars can form only in structures that are suitably dense can be parameterized by the threshold mass $M_{\rm min}$. With this definition, the baryon accretion rate $a_{\rm b}(t)$, which accounts for the increase in the fraction of baryons in structures, is given by (see \citeauthor{pereira2010} \citeyear{pereira2010} and references therein)

\begin{equation}
a_{\rm b}(t) = \left(\Omega_{\rm b}\rho_{\rm c}\right)\frac{df_{\rm b}}{dt},\label{abaryon}
\end{equation}

\noindent where $\rho_{\rm c}=3H_{0}^{2}/8\pi G$ is the critical density of the universe ($H_{0}=100\,h\,{\rm km}\,{\rm s}^{-1}\,{\rm Mpc}^{-1}$ is the value of the Hubble parameter at the current time).

To complete the cosmological part of CSFR, we need to normalize the power spectrum. We often choose to express this normalization in terms of a parameter called $\sigma_{8}$, which represents the value of $\sigma(M)$ at $z =0$ within a sphere of radius $R= 8h^{-1}{\rm Mpc}$. Following \cite{ade2016} we can find $\sigma_{8} = 0.830\pm 0.015$. Once we have followed these steps, we will have the cosmological part of the CSFR well characterized. In particular, the set of Equations described above synthesizes the fundamental basis for the theory of cosmological perturbations, which consequently leads to the formation of large-scale structures of the universe. As discussed by PM, the CSFR can then be constructed from this scenario simply by incorporating the laws of Schmidt and Salpeter. To do this, we should remember that the star formation for a galactic-like system is determined by the interplay between incorporation of baryons into collapsed objects (stars, stellar remnants, and smaller objects) and return of baryons into a diffuse state (such as gaseous clouds and the intercloud medium of the system). 

The second process can be two-fold: (a) mass return from stars to the interstellar medium (ISM) through, for example, stellar winds, and supernovae, which happens at the local level; and (b) net global infall of baryons from outside of the system. The former process is a well-known and firmly established part of the standard stellar evolution lore, and although details of mass-loss in a particular stellar type may still be controversial, there is nothing controversial in the basic physics of this process. Thus, we consider the baryon accretion rate $a_{\rm b}(t)$, described by Eq. (\ref{abaryon}), as an infall term that supplies the reservoir represented by the halos. Therefore, the number of stars formed by unity of volume, mass, and time is given by:

\begin{equation}
\frac{d^{3}N}{dVdmdt} = \Phi (m) \Psi(t),\label{d3n}
\end{equation}

\noindent where $\Phi (m)$ is the IMF that gives the distribution function of stellar masses, and $\Psi (t)$ is the star formation rate. See that $\Psi (t)$ is assumed to be independent of mass while $\Phi (m)$ is assumed to be independent of time. Using the Schmidt law \citep{schmidt1959,schmidt1963} for $\Psi (t)$, we have

\begin{equation}
\frac{d^{2}M_{\star}}{dVdt} = \Psi (t) = k[\rho_{\rm g}(t)]^{\alpha},\label{sclaw}
\end{equation}

\noindent where $k$ is a constant that will be identified later, $\rho_{\rm g}$ is the local gas density, and $\alpha=1$. See that (\ref{sclaw}) shows that stars are formed by the gas contained in the halos.

On the other hand, we assume that the IMF follows the \cite{salpeter1959} form

\begin{equation}
\Phi(m) = A m^{-(1+x)}\label{imf1},
\end{equation}

\noindent where $x$ is the Salpeter exponent and $A$ is a normalization factor.

The constant $A$ is determined by the condition that all stars are formed into the mass range $[m_{\rm inf},m_{\rm sup}]$. That is,

\begin{equation}
\int_{m_{\rm inf}}^{m_{\rm sup}}Am^{-(1+x)}mdm = 1,\label{norimf}
\end{equation}

\noindent and we consider $m_{\rm inf}=0.1\,M_{\odot}$ and $m_{\rm sup}=140\,M_{\odot}$ as limits in (\ref{norimf}).

The mass ejected from stars can be determined by 

\begin{equation}
\frac{d^{2} M_{\rm ej}}{dVdt} = \int_{m(t)}^{\rm m_{sup}}{(m-m_{\rm r})\Phi(m)\Psi(t-\tau_{m})dm},\label{mej1}
\end{equation}

\noindent where the lower limit of the integral, $m(t)$, corresponds to the stellar mass whose lifetime is equal to $t$. The term $m_{\rm r}$ represents the mass of the remnant, which depends on the progenitor mass. The star formation rate is taken at the retarded time $(t-\tau_{\rm m})$, where $\tau_{\rm m}$ is the lifetime of a star of mass $m$ which can be calculated by means of \citep{s6,c4}

\begin{equation}
 \log_{10}(\tau_{\rm m})=10.0-3.6\,\log_{10}\left(\frac{M}{M_{\odot}}\right) +\left[ \log_{10}
\left( \frac{M}{M_{\odot}}\right) \right]^{2},
\end{equation}

\noindent where $\tau_{\rm m}$ is the stellar lifetime given in years.

The mass of the remnant, $m_{\rm r}$, in Eq. (\ref{mej1}) is determined using the following assumptions:

\noindent a) Stars with $m < 1\ {M}_{\odot}$ have a high lifetime, so they do not contribute to $M_{\rm ej}$;

\noindent b) Stars with $1\ {M}_{\odot} \leq m\leq 8\ {M}_{\odot}$ after evolving off the main sequence left carbon--oxygen white dwarfs as remnants, where 

\begin{equation}
m_{\rm r} = 0.1156\ m +0.4551;
\end{equation}

\noindent c) Stars in the range $8\ {M}_{\odot} < m\leq 10\ {M}_{\odot}$ after evolving off the main sequence left oxygen-neon-magnesium white dwarfs with $m_{\rm r} = 1.35\ {M}_{\odot}$;

\noindent d) Stars with $10\ {M}_{\odot} < m< 40\ {M}_{\odot}$ explode as supernovae, leaving neutron stars as remnants ($m_{\rm r}=1.4\ {M}_{\odot}$);

\noindent e) Stars with $40\ {M}_{\odot} \leq m\leq 140\ {M}_{\odot}$ produce black hole remnants. In this case, we consider (see \citealt{h1})

\begin{equation}
m_{\rm r}=\frac{13}{24}(m-20\ {M}_{\sun}).\label{mr}
\end{equation} 

We can then write an equation governing the total gas density ($\rho_{\rm g}$) in the halos. Namely,

\begin{equation}
 \dot\rho_{\rm g}=-\frac{d^{2}M_{\star}}{dVdt}+\frac{d^{2}M_{\rm ej}}{dVdt}+a_{\rm b}(t)\label{rhogas},
\end{equation}

\noindent where $a_{\rm b}(t)$, Eq. (\ref{abaryon}), gives the rate at which the halos accrete baryonic (gas) mass.

Numerical integration of (\ref{rhogas}) produces the function $\rho_{\rm g}(t)$ at each time $t$ (or redshift $z$). Once obtained $\rho_{\rm g}(t)$, we return to Eq. (\ref{sclaw}) in order to obtain the ``CSFR" $\Psi(t)$. Just replacing $\Psi(t)$ by $\dot\rho_{\star}$, we can write

\begin{equation}
\dot\rho_{\star}=k\rho_{\rm g}\label{csfr},
\end{equation}

\noindent where the constant $k$ represents the inverse of the timescale for star formation. Namely, $k=1/\tau_{\rm s}$.

The CSFR, as presented in Equation (\ref{csfr}), is not yet in its final form; it is necessary to normalize it. This can be done by introducing a factor $\langle\varepsilon\rangle$ that causes the CSFR to take the value $\dot\rho_{\star}=0.016\,M_{\odot}\,{\rm yr}^{-1}\,{\rm Mpc}^{-3}$ at $z=0$ (see \citeauthor{pereira2010} \citeyear{pereira2010}; \citeauthor{pereira2011} \citeyear{pereira2011}). This value produces good agreement with both the present value of the CSFR derived by \cite{sprher}, who employed hydrodynamic simulations of structure formation, and the observational data taken from \cite{h2,h3}. The normalization is also related to the fact that not all gas captured by halos will be transformed into stars. Thus, the final form of CSFR is

\begin{equation}
\dot\rho_{\star}(z)=\, \langle\varepsilon\rangle\frac{\rho_{\rm g}}{\tau_{\rm s}}\label{csfr2},
\end{equation}

\noindent where $\langle\varepsilon\rangle$ is the efficiency for star formation, which also acts as a normalization factor for the CSFR at $z=0$.

There are two strong consequences associated with Equation (\ref{csfr2}). The first is that the CSFR is a weighted average that also depends on the masses of all halos capable of collapsing in a given redshift. At high redshift ($z\sim 20$), we have predominantly the formation of halos with masses close to $10^ {6}-10^{7}M_{\odot}$, while at low redshifts we find, in addition to a large number of halos of low masses, halos with masses comparable to galaxies. The second point, as commented above, is associated with the fact that not all gas is used to form stars. This can be represented as

\begin{equation}
\langle\varepsilon\rangle = \frac{\rho_{\rm mol}(z)}{\rho_{\rm g}(z)}\label{efficiency},
\end{equation}

\noindent where $\rho_{\rm mol}$  is the fraction of the total gas directly used to form stars. As a consequence of this, $\langle\varepsilon\rangle$ is a function of the redshift. It is important to note that the definition for the efficiency of star formation through Equation (\ref{efficiency}) is equivalent to the usual definition $\langle\varepsilon\rangle=\rho_{\star}/\rho_{\rm g}$, because $\rho_{\star}\equiv\dot\rho_{\star}\tau_{\rm s}$ in our model.

Table \ref{initialp} summarizes the parameters used to obtain the CSFR. As discussed in PM, the best agreement with the observational data is achieved with $x = 1.35$ (Salpeter exponent) and $\tau_{\rm s} = 2.0\,{\rm Gyr}$, which is the characteristic timescale for star formation. The behavior of this model can be seen in Figure 1. See that $\langle\varepsilon\rangle = 0.021$ at $z = 0$, in order to obtain $\dot\rho_{\star}(z=0)=0.016\,M_{\odot}\,{\rm yr}^{-1}\,{\rm Mpc}^{-3}$. The evolution of $\langle\varepsilon\rangle$ with redshift can be seen in Figure 2. Note that star formation efficiency is high $\langle\varepsilon\rangle = 0.32$ at high redshifts ($z\sim 20$), reaching 0.021 at $z = 0$.

\begin{table}[!ht]
\begin{center}
\caption{Parameters of the CSFR}
\centering
\begin{tabular}{cccccccccc}
\hline
\hline 
$\Omega_{\rm m}$ & $\Omega_{\rm b}$ & $\Omega_{\Lambda}$ & $h$ & $z$ & $n_{\rm p}$ & $\sigma_{8}$ & $\tau_{\rm s}(\rm Gyr)$ & $M_{\rm min}(M_{\odot})$ & $x$ \\ 
\hline 
0.279& 0.0463 & 0.721 & 0.7 & 20 & 0.97 & 0.84 & 2.0 & $10^{6}$ & 1.35 \\ 
\hline
\end{tabular} 
\label{initialp}
\end{center}
{\bf Note.} $\Omega_{\rm m}$ corresponds to the total matter (baryonic plus dark matter) density parameter; $\Omega_{\rm b}$ is the baryonic density parameter; $\Omega_{\Lambda}$ is the density parameter associated with dark energy (cosmological constant); $h$ is the Hubble constant written as $H_{0}=100\,h\,{\rm km}\,{\rm s}^{-1}\,{\rm Mpc}^{-1}$; $z$ is the redshift at which star formation begins; $n_{\rm p}$ is the exponent of the primordial power spectrum; $\sigma_{8}$ is the normalization of the power spectrum, in other words $\sigma(M,0)$; $\tau_{\rm s}$ is the timescale for star formation; $M_{\rm min}$ corresponds to the lowest mass a halo of dark matter must have to detach from the expansion of the universe, to collapse and to virialize (it is approximately equal to the Jeans mass at recombination); $x$ is the exponent of the IMF.
\end{table}

\begin{figure}[ht!]
\epsscale{0.51}
\plotone{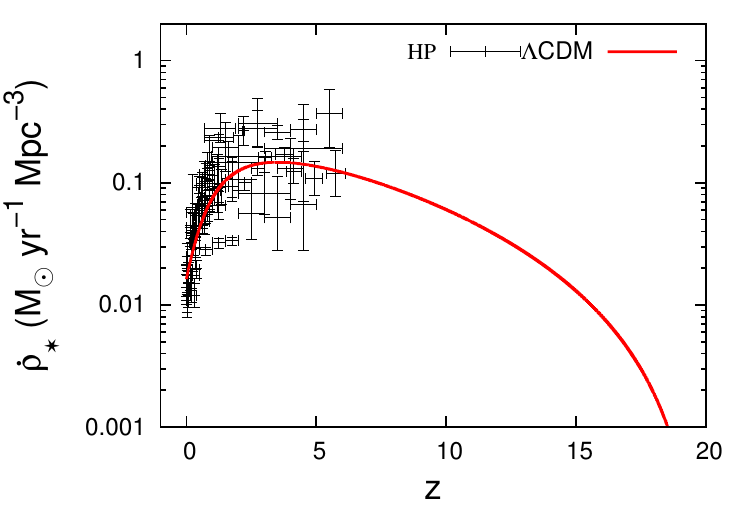}
\caption{The evolution of the CSFR with redshift derived for the hierarchical structure formation scenario (standard $\Lambda$CDM cosmological model). At redshift 3.5, the CSFR achieves maximum value ($\dot\rho_{\star}=0.147\,M_{\odot}\,{\rm yr}^{-1}\,{\rm Mpc}^{-3}$). The observational points (HP) are taken from \citet{h2,h3}.\label{fig:csfr}}
\end{figure}

\begin{figure}[ht!]
\epsscale{0.51}
\plotone{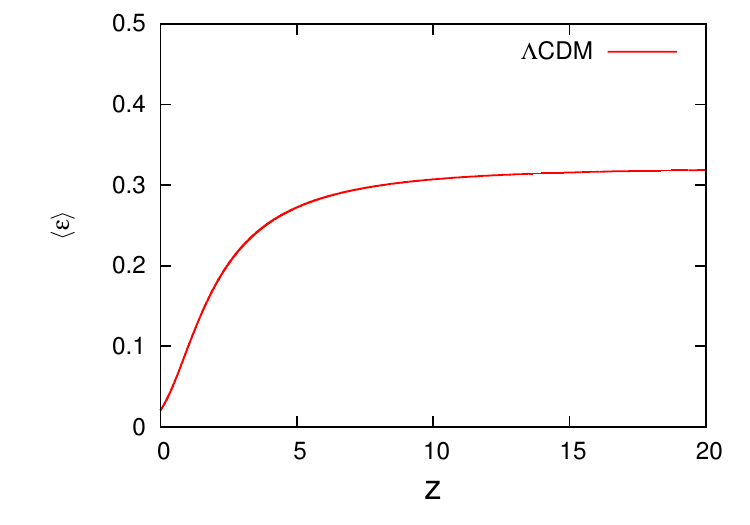}
\caption{The star formation efficiency as a function of the redshift. The determination of $\langle\varepsilon\rangle$ is done through Equation (\ref{efficiency}). The efficiency is almost constant within the range $\sim 3.5-20$ in redshift, with an average value close to $\langle\varepsilon\rangle \approx 0.32$. For $z\lesssim 3.5$, the efficiency rapidly decreases, reaching $0.021$ at $z = 0$.\label{fig:e_csfr}}
\end{figure}

\subsection{Local Star Formation Rate--SFR} \label{subsec:sfr}

The ISM is a gaseous medium intrinsically connected with the life cycle of stars. The ISM provides us with a very rich physics through the interactions of stellar winds, supernova explosions, jets associated with proto-stellar systems, among others. This rich interaction causes the ISM to be a complex and filamentous structure that consequently produces turbulent movements in the gas which, in turn, regulate the star formation (see \citeauthor{federrath2017} \citeyear{federrath2017}, for a review of turbulence drivers). Since the work of \cite{krumholz2005}, it has been discussed in the literature that the small values for the star formation efficiencies, as highlighted in the section \ref{sec:intro}, could be associated with the supersonic turbulent motions of the gas in the star formation regions. In particular, turbulence is a self-similar process that can carry energy from the large scale to the small. Thus, turbulence could provide the necessary support to retard the gravitational collapse of the gas so that the star formation would result from the gravo-turbulent fragmentation of the molecular clouds (\citeauthor{maclow2004} \citeyear{maclow2004}).

The presence of turbulent motions with high Mach numbers could create broad (log-normal) distributions for the gas density. In this way, when we analyze star formation at the local level (i.e., for redshifts $z\approx 0$), it is common to use the so-called density probability distribution functions (PDFs) of the column gas density, as well as the volume gas density, as common tools for studying these star-forming regions. For a purely isothermal gas, the PDF has the form (see, e.g., \citeauthor{vazquez94} \citeyear{vazquez94}; \citeauthor{passot98} \citeyear{passot98}; \citeauthor{vazquez2003} \citeyear{vazquez2003})

\begin{equation}
p(s) = \frac{1}{\sqrt{2\pi\sigma_{\rm s}^{2}}}\,{\rm exp}\left[-\frac{(s-s_{0})^{2}}{2\sigma_{\rm s}^{2}}\right],\label{lognormal}
\end{equation}

\noindent where $s={\rm ln}(\rho_{\rm g}/\rho_{0})$ is the logarithmic density contrast, $\sigma_{\rm s}^{2}$ is the density variance, and $s_{0}=-\sigma_{\rm s}^{2}/2$  is the mean value that is related to the density variance due to mass conservation. As pointed out by \cite{hopkins2013b}, if the gas can be considered isothermal, then Equation (\ref{lognormal}) may represent density fluctuations in both subsonic and supersonic regimes. 

The density variance is a function of the root-mean-squared (rms) Mach number ($\mathcal{M}$), and is given by

\begin{equation}
\sigma_{\rm s}^{2} = {\rm ln}\left(1+b^{2}\mathcal{M}^{2}\frac{\beta}{1+\beta}\right).\label{variance}
\end{equation}

The coefficient $b$ is known as the turbulence driving parameter; it is related to the mixture mode induced by the turbulent forcing mechanism. The value $b = 1/3$ corresponds to the purely solenoidal driving, while $b = 1$ is associated with the purely compressive driving (see, e.g., \citeauthor{federrath2008} \citeyear{federrath2008}; \citeauthor{federrath2010} \citeyear{federrath2010}). The $\beta$ parameter represents the ratio between the thermal and magnetic pressures (see, e.g., \citeauthor{padoan2011} \citeyear{padoan2011}; \citeauthor{federrath2012} \citeyear{federrath2012}; \citeauthor{molina2012} \citeyear{molina2012})\footnote{As discussed in \cite{federrath2012}, the definition of the $\beta$ parameter as done by \cite{padoan2011} is slightly different from that considered in Equation (\ref{variance}).}. In the case of no density correlation of the magnetic field, we have $B\propto \rho^{0}$ and so $\beta \rightarrow \infty$ producing (see, e.g., \citeauthor{padoan1997} \citeyear{padoan1997}; \citeauthor{passot98} \citeyear{passot98}; \citeauthor{price2011} \citeyear{price2011})

\begin{equation}
\sigma_{\rm s}^{2} = {\rm ln}\left(1+b^{2}\mathcal{M}^{2}\right).\label{variance2}
\end{equation}

Once we have the PDF, it is enough to integrate it from a certain threshold to infinity to obtain the SFR. Integration can be weighted by $\rho_{\rm g}/t_{\rm ff}(\rho_{\rm g})$, where $t_{\rm ff}(\rho_{\rm g})=(3\pi/32G\rho_{\rm g})^{1/2}$ is the freefall time. The result is then

\begin{equation}
{\rm SFR}\sim \int_{\rho_{\rm crit}}^{\infty} \frac{\rho_{\rm g}}{t_{\rm ff}(\rho_{\rm g})}p(\rho_{\rm g})d\rho_{\rm g}.\label{sfr1}
\end{equation}

Equation (\ref{sfr1}) is known as the ``multi-freefall model" of the SFR. It can be written in terms of the logarithmic density $s\equiv {\rm ln}(\rho_{\rm g}/\rho_{0})$, producing

\begin{equation}
{\rm SFR}\sim \int_{s_{\rm crit}}^{\infty} {\rm exp}\left(\frac{3}{2}s\right)p(s)ds.\label{multifreefall}
\end{equation}

By plugging Equation (\ref{lognormal}) into (\ref{multifreefall}), it is possible to analytically solve the integral, which results in

\begin{equation}
{\rm SFR}\sim \frac{1}{2}{\rm exp}\left(\frac{3}{8}\sigma_{\rm s}^{2}\right)\left[1+{\rm erf}\left(\frac{\sigma_{\rm s}^{2}-s_{\rm crit}}{\sqrt{2\sigma_{\rm s}^{2}}}\right)\right].\label{sfr2}
\end{equation}

The model presented above has been used by different authors to characterize the SFR from the PDF of the density fluctuations induced in the clouds by the turbulence. These models are primarily characterized by the definition of a density threshold usually represented as $s_{\rm crit}\sim {\rm ln}(\alpha_{\rm vir}\mathcal{M}^{2})$, where $\alpha_{\rm vir}$ is the so-called virial parameter and $\mathcal{M}$ is the Mach number (see \citeauthor{federrath2012} \citeyear{federrath2012}, for a derivation of Eqs. \ref{sfr1}-\ref{sfr2}). The virial parameter is a measurement of the level of turbulence versus gravitational energy of an object. Thus, it is given by $\alpha_{\rm vir}=2E_{\rm k}/E_{\rm p}$, where $E_{\rm k}$ is the kinetic energy and $E_{\rm p}$ is the gravitational potential energy \citep{bertoldi92}. From this definition, we can write $\alpha_{\rm vir}=5\sigma_{0}^2 R/GM$, where $M$, $R$, and $\sigma_{0}$ are, respectively, the mass, radius, and rms velocity within the object (we are using ``object" as a synonym for both molecular clouds and molecular clumps).

Note that, to quantify $\alpha_{\rm vir}$, we must define a region where the parameters $\sigma_{0}$ and $M$ can be estimated. The virial parameter is only suitable for the clouds that have well-defined structures \citep{guang2015}. However, the morphology of molecular clouds is, in general, quite complicated. In many cases, it is not trivial to separate individual clouds from the surrounding environment. Indeed, the fact that clouds are neither isolated, nor spherical, nor of uniform density can lead to an order of magnitude difference in virial parameter (see \citeauthor{federrath2012} \citeyear{federrath2012}). Moreover, as the clouds are observed projected on the plane of the sky, the morphology of these objects can be biased by projection effects (see, e.g., \citeauthor{pichardo2000} \citeyear{pichardo2000}; \citeauthor{dib2006} \citeyear{dib2006}; \citeauthor{shetty2010} \citeyear{shetty2010}; \citeauthor{beaumont2013} \citeyear{beaumont2013}). Thus, there is a large uncertainty concerning the estimated virial parameters in the literature (see, e.g., \citeauthor{rosolowsky2007} \citeyear{rosolowsky2007}; \citeauthor{hernandez2015} \citeyear{hernandez2015}). In particular, \cite{padoan2016,padoan2017} have analyzed the SFR as a function of the cloud parameters, obtaining values within the range $\sim 0.5-25$ for the virial parameter. On the other hand, \cite{hennebelle2011} have preferred not to set a threshold for star formation. In contrast, these authors consider that SFR continuously increases with gas density, thus producing two different characteristic regimes.

A step further in the SFR study comes with the so-called polytropic turbulence models. As pointed out by \cite{federrath2015}, some important works developed mainly in the last ten years have shown that the PDF tends to deviate from the lognormal form given by Equation (\ref{lognormal}) if the gas is non-isothermal (see \citeauthor{federrath2015} \citeyear{federrath2015} and references therein; \citeauthor{nolan2015} \citeyear{nolan2015}). A physically well-motivated functional form for a non-isothermal PDF was suggested by \cite{hopkins2013a}. As pointed out by the author, the proposed function is considerably good when compared to data on a large Mach number range, and variance in numerical simulations. In particular, as shown in H13 and FB15, the fit for this PDF is

\begin{eqnarray}\label{hksfr}
	p_{\rm hk}(s) = I_{1}\left(2\sqrt{\lambda\omega(s)}\right)\,{\rm exp}\left[-(\lambda + \omega(s))\right]\sqrt{\frac{\lambda}{\theta^{2}\omega(s)}}\nonumber \\
    \lambda\equiv \frac{\sigma_{\rm s,V}^{2}}{2\theta^{2}};\quad \omega(s)\equiv \frac{\lambda}{(1+\theta)}-\frac{s}{\theta}\quad (\omega(s)\geq 0),
\end{eqnarray}

\noindent where $I_{1}(x)$ is the modified Bessel function of the first kind. The parameter $\sigma_{\rm s,V}$ is the volume-weighted standard deviation of the logarithmic density fluctuations, while $\theta$ is the intermittency parameter. In the zero-intermittency limit ($\theta \rightarrow 0$), Equation (\ref{hksfr}) becomes the lognormal distribution from Equation (\ref{lognormal}).

In order to obtain the SFR from non-isothermal PDF (Equation \ref{hksfr}), it is necessary to adequately characterize $\sigma_{\rm s}$, because the form given in Equation (\ref{variance2}) applies only to the isothermal case. There are two different ways to do this. The first is to follow FB15, who use the Rankine-Hugoniot conditions to obtain the following equation for the density contrast $x\equiv \rho_{\rm g}/\rho_{0}$

\begin{equation}\label{fb2015}
x^{\Gamma}+\Gamma b^{2}\mathcal{M}^{2}\left(\frac{1}{x}-1\right)-1=0,
\end{equation}

\noindent where $\Gamma$ is the polytropic index.

As pointed out by FB15, solving the transcendental Equation (\ref{fb2015}), we obtain the variable $s={\rm ln}(\rho_{\rm g}/\rho_{0})$ and its logarithmic density variance, which is given by

\begin{equation}
\sigma_{\rm s}^{2} \simeq {\rm ln}\left(1+\frac{\rho_{\rm g}}{\rho_{0}}\right),\label{sigma_int}
\end{equation}

\noindent where for $\Gamma = 1$, the non-trivial solution of Equation (\ref{fb2015}) yields $x\equiv \rho_{\rm g}/\rho_{0} = b^{2} \mathcal{M}^{2}$, such that we retrieve $\sigma_{\rm s}$ for the isothermal case, as given by Equation (\ref{variance2}).

The connection between $\sigma_{\rm s,V}$ and $\sigma_{\rm s}$ is made through the intermittency parameter ($\theta$), which in turn is related to $\Gamma$ ($\Gamma\neq 1$) by a power law (see the discussion on these points presented in FB15). Thus,

\begin{equation}\label{theta}
\theta = 0.035b\mathcal{M}\Gamma^{2},
\end{equation}

\noindent which in turn produces

\begin{equation}\label{iso1}
\sigma_{\rm s,V}^{2} = \sigma_{\rm s}^{2}\left(1+\theta\right)^{3/2}.
\end{equation}

The second way to characterize $\sigma_{\rm s}$ is presented by \cite{nolan2015} specifically for adiabatic turbulence. In particular, these authors used high-resolution hydrodynamic simulations to investigate the relationship between $\sigma_{\rm s}$ and $\mathcal{M}$ in both isothermal and non-isothermal regimes. Their main result is a new relationship between density variance and Mach number, given by

\begin{equation}\label{iso2}
\sigma_{\rm s}^{2} = {\rm ln}\left[1+b^{2}\mathcal{M}^{(5\gamma + 1)/3}\right],
\end{equation}

\noindent for $b\mathcal{M}\lesssim 1$, and

\begin{equation}\label{iso3}
\sigma_{\rm s}^{2} = {\rm ln}\left[1+\frac{\left(\gamma+1\right)b^{2}\mathcal{M}^{2}}{\left(\gamma-1\right)b^{2}\mathcal{M}^{2}+2}\right],
\end{equation}

\noindent for $b\mathcal{M} > 1$, where $\gamma$ is the adiabatic index.

\cite{nolan2015} conclude that, to study adiabatic turbulence, these relationships can introduce important corrections, especially if the gas is non-isothermal ($\gamma\neq 1$). In this paper, however, we will strictly follow the formalism presented by FB15. Because we have  $\Gamma$, $\mathcal{M}$, $b$ beside $\sigma_{\rm s,V}$, then it is possible to use the Hopkins PDF (\citeauthor{hopkins2013a} \citeyear{hopkins2013a}) to obtain the SFR as (see, in particular, \citeauthor{federrath2015} \citeyear{federrath2015}).

\begin{equation}\label{sfrhopkins}
{\rm SFR} \sim \int_{s_{\rm crit}}^{\infty}{\rm exp}\left(\frac{3}{2}s\right)p_{\rm hk}(s)ds.
\end{equation}

Equation (\ref{sfrhopkins}) holds for the $\Gamma \neq 1$ cases. In order to make it an equality, we must define the right-hand multiplicative factor: it can be in the form $\Sigma_{\rm gas}/t_{\rm dep}$ if we wish to express the SFR in units of $M_{\odot}\,{\rm kpc^{-2}\,yr^{-1}}$, or of the form $\rho_{\rm gas}/\tau_{\rm s}$ if we wish to express it in units of $M_{\odot}\,{\rm yr}^{-1}\,{\rm Mpc}^{-3}$. Because we wish to discuss the possible complementarity between global star formation (CSFR) and local star formation (SFR), it is more appropriate to take the latter form by rewriting the Equation (\ref{sfrhopkins}) as

\begin{equation}\label{sfrfinal}
\dot\rho_{\rm SFR} = \frac{\rho_{\rm g}}{\tau_{\rm s}} \int_{s_{\rm crit}}^{\infty}{\rm exp}\left(\frac{3}{2}s\right)p_{\rm hk}(s)ds,
\end{equation}

\noindent where $\rho_{\rm g}$ is the gas density and $\tau_{\rm s}$ is the timescale for star formation.

Before closing this section, it is important to discuss one more aspect associated with the SFR, which is the so-called Larson's law\footnote{We are considering in this article just the Larson relation, which is known in literature as Larson's first law. The so-called second law shows the relationship between the velocity dispersion and the mass of the cloud. The third shows that the size of the cloud is inversely proportional to the density.}. In a seminal paper, \cite{larson81} proposed that the protostellar cores are originated by turbulent supersonic compression, which in turn causes gravity to become dominant only in the denser regions (which generally possess subsonic characteristics).

Larson used measures of the velocity dispersion, $V_{\rm rms}$, of molecular clouds showing that, on a scale of $0.1 < R < 100\,{\rm pc}$, this is given by $V_{\rm rms} \propto R^{0.38}$. On the other hand, \cite{solomon87} have found a slightly different power index $\sim 0.5$ (see also \citeauthor{federrath2013b} \citeyear{federrath2013b} and references therein). More recently, \cite{kritsuk2013} have reviewed the origin of Larson's law using recent observational measurements as well as numerical simulations of the ISM. These authors argue that Larson's relations on scales of $0.1-50\,{\rm pc}$ can be interpreted as supersonic turbulence fed by the large-scale kinetic energy injection. However, most likely there are multiple injection mechanisms on multiple scales acting together in a complex way in the ISM (see, e.g., \citeauthor{federrath2017} \citeyear{federrath2017}). Thus, a single power law may result in simplification of the problem.

Following the notation of \cite{hennebelle2008,hennebelle2009}, we will represent Larson's first law as

\begin{equation}\label{larsonlaw}
\langle {V_{\rm rms}^ {2}} \rangle = V_{0}^{2}\left( \frac{R}{1\,{\rm pc}}\right)^{2\eta},
\end{equation}

\noindent where $V_{0}\simeq 1\,{\rm km\,s^{-1}}$, and $\eta \simeq 0.4-0.5$. Once we have reviewed the bases of CSFR and SFR, we are in a position to explore the complementarity between these star formation rates.

\section{Model Unifying CSFR with SFR---Results}\label{results}

\subsection{How the SFR can Mimic the Evolution of the CSFR}\label{mimic}

Our {\it ansatz} considers that Equations (\ref{csfr2}) and (\ref{sfrfinal}) represent the same physics, which can be applied on both the cosmological and local galactic scales. In this way, we propose that the following equality is valid:

\begin{equation}\label{igual1}
\frac{\dot\rho_{\star}(z)}{\langle\varepsilon\rangle} = \dot\rho_{\rm SFR},
\end{equation}

\noindent or explicitly

\begin{equation}\label{igual2}
\frac{\dot\rho_{\star}(z)}{\langle\varepsilon\rangle} = \frac{\rho_{\rm g}}{\tau_{\rm s}} \int_{s_{\rm crit}}^{\infty}{\rm exp}\left(\frac{3}{2}s\right)p_{\rm hk}(s)ds.
\end{equation}

Looking from the viewpoint of CSFR, the difference between Equations (\ref{csfr2}) and (\ref{igual2}) is associated with the inclusion of Hopkins' PDF (which, as shown by FB15, can be linked with $\Gamma\neq 1$) or isothermal PDF (if $\Gamma=1$). In order to maintain consistency with the results derived by PM for the CSFR and reviewed in section \ref{subsec:csfr}, the integral in Equation (\ref{igual2}) must be equal to 1 at redshift $z=0$ to reproduce the Equation (\ref{csfr2}). We then set the parameters $\Gamma$, $b$, and $s_ {\rm crit}$ in an attempt to solve the integral as a function of the Mach number. That is, we look for the value of $\mathcal{M}$ that allows to recover the value of the CSFR at $z = 0$. The second point considered in our model is to verify if the PDF used to characterize the SFR may or may not ``mimic" the CSFR. That is, we take the redshift variation of the parameters $\dot\rho_{\star}$ and $\langle\varepsilon\rangle$ on the left side of Equation (\ref{igual2}) and we maintain $\rho_{\rm g}$, on the right side of the equation, ``frozen" for its value at $z = 0$. Then, we determine the values of $\mathcal{M}$ that satisfy the equality of this equation for each redshift value that composes the CSFR. For all models, we set the parameter $b = 0.4$ (see \citeauthor{federrath2015} \citeyear{federrath2015}), while the characteristic timescale ($\tau_{\rm s}$) for star formation is $2\,{\rm Gyr}$.

\begin{figure}[ht!]
\epsscale{1.15}
\plottwo{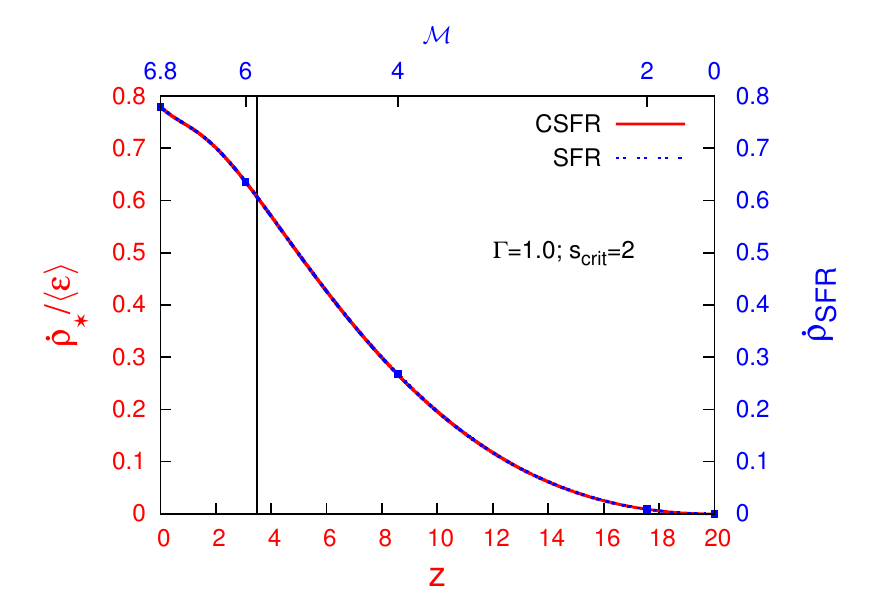}{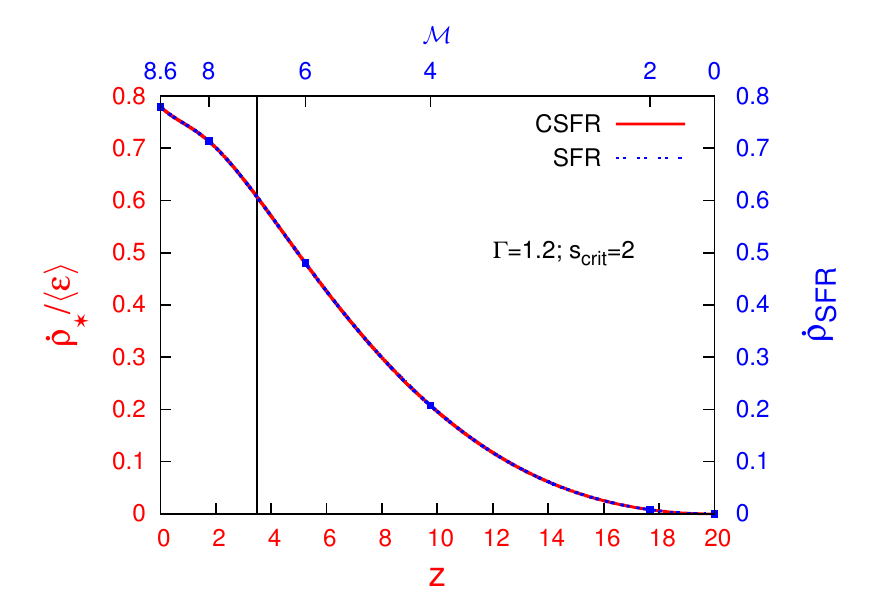}
\plottwo{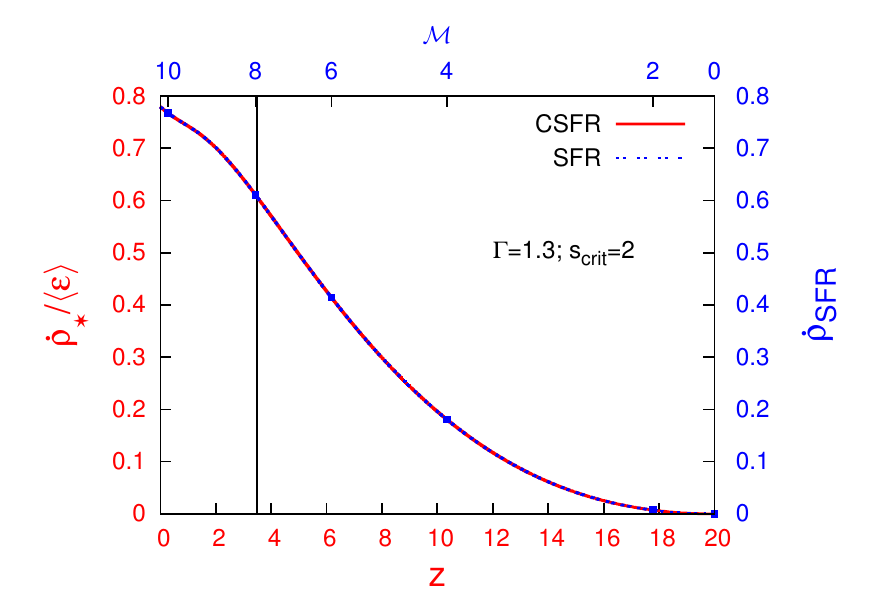}{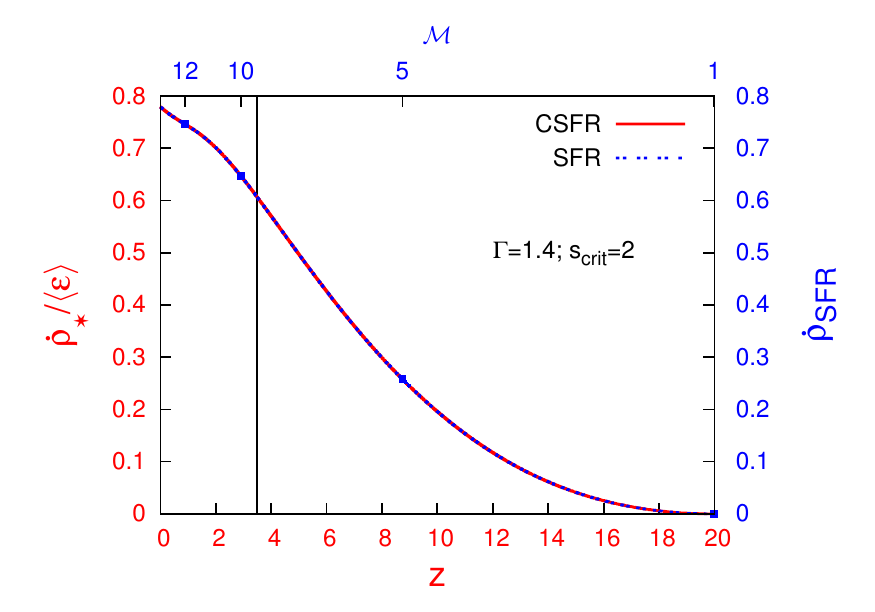}
\caption{The solution of the Equation (\ref{igual2}) with $s_{\rm crit}=2$, for different values of $\Gamma$. The evolution of the CSFR, weighted by the average efficiency, as a function of the redshift, is presented in the $y1-x1$ axes (in red). The axes $y2-x2$ (in blue) show the SFR obtained with the frozen value of $\rho_{\rm g}$ at $z = 0$, and looking for the value of the Mach number that satisfies the equality of this equation. This allows us to relate the redshift, provided by the CSFR, to the Mach number, provided by the SFR. With this analysis, it is possible to infer the role of turbulence associated with the formation of stars at high redshifts which, in turn, allows to generate the large-scale structures observed in the universe.\label{fig4acde}}
\end{figure}

\begin{figure}[ht!]
\epsscale{1.15}
\plottwo{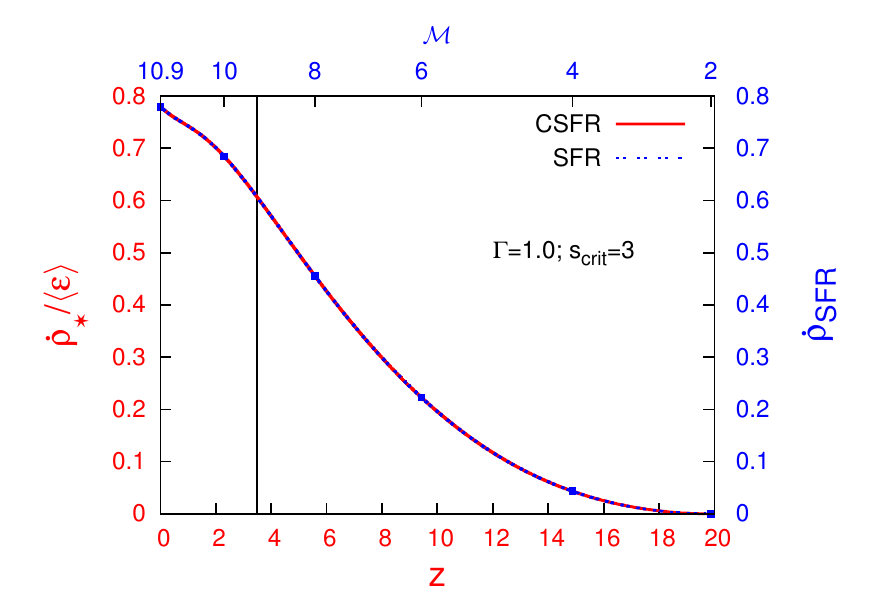}{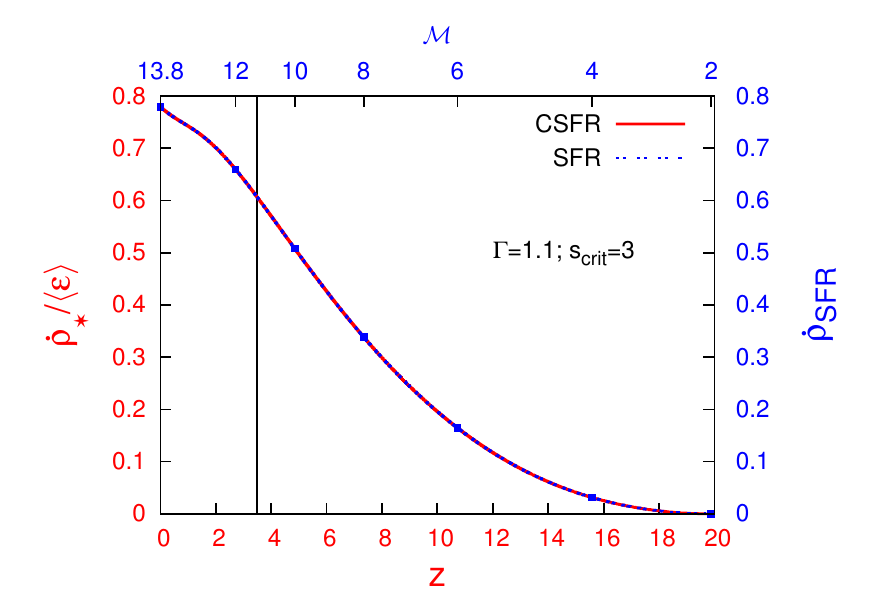}
\plottwo{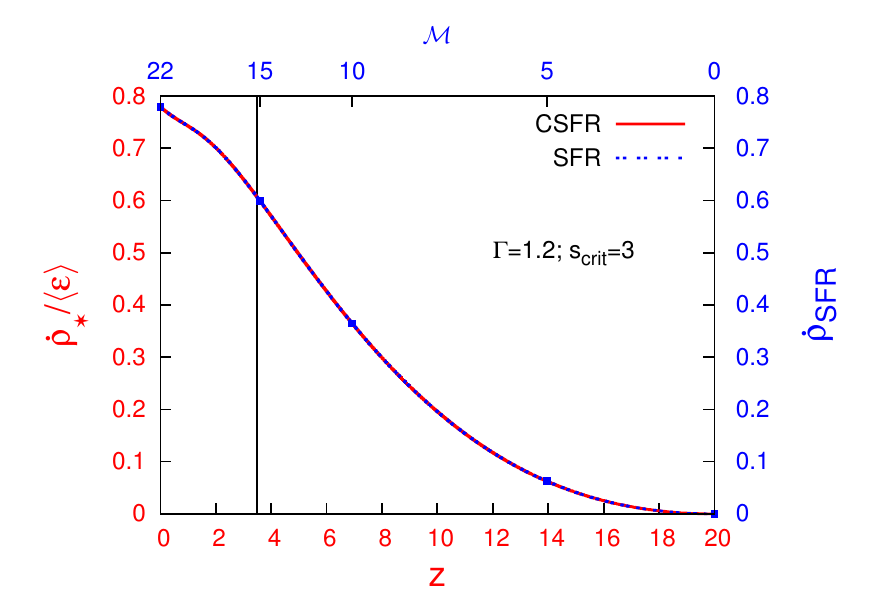}{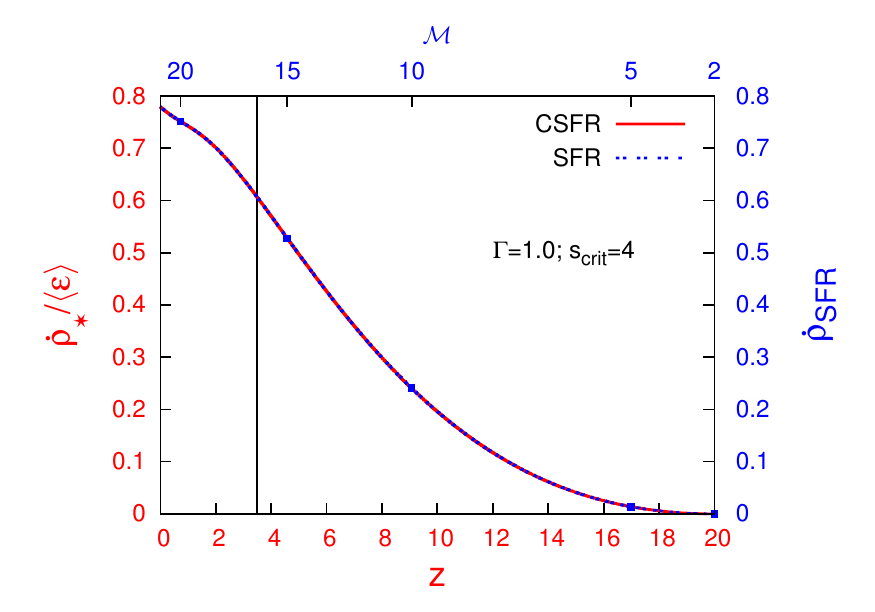}
\caption{The solution of the Equation (\ref{igual2}) with $s_{\rm crit}=3$ and $4$ considering different values of $\Gamma$. The results are similar to those observed in Figure \ref{fig4acde}, although the $\mathcal{M}$ versus $z$ relationship changes with $s_{\rm crit}$ when we look at the same polytropic index. \label{fig5abcd}}
\end{figure}

Figure \ref{fig4acde} shows the results obtained for $s_{\rm crit} = 2$ and different values of $\Gamma$. The black vertical line marks the redshift ($z = 3.5$) where the CSFR reaches its maximum value. Depending on the value of $\Gamma$, the Mach number reaches values within the range $\sim 6-9$ at redshift $z = 3.5$. With the increment of redshift ($z\rightarrow 20$), it is possible to verify that $\mathcal{M}$ decreases. In particular, at $z=0$ we have $\mathcal{M}$ within the interval $\sim 7-13$, depending on the particular value of $\Gamma$.

Figure \ref{fig5abcd} exhibits the results for $s_{\rm crit} = 3$ and $4$. It can be seen that the results are similar to those obtained for the case $s_{\rm crit} = 2$. Note, however, that the relation $\mathcal{M}$ versus $z$ changes with the increase of $s_{\rm crit}$. Considering $s_{\rm crit} = 3$, the peak of the CSFR corresponds to $\mathcal{M}$ within the range $\sim 9-15$, while at $z = 0$ the Mach number lies in the range $\sim 11-22$. The Table \ref{resumo} presents, for the nine different models generated in our analysis, the values reached for the Mach number at both the CSFR peak ($z = 3.5$) and at $z = 0$.

From Figures \ref{fig4acde} and \ref{fig5abcd}, we can immediately verify that indeed the SFR can, through the Mach number, ``mimic" the evolution of the CSFR from redshift $\sim 20$ up to the present, with the two curves having excellent agreement. However, to quantify this agreement between the CSFR and the SFR, we divide the redshift interval $0-20$ into 12,000 linearly spaced points, inferring the degree of deviation ($D$) from the equality represented by Equation (\ref{igual2}). Table \ref{err} shows the result of this analysis. In particular, we evaluate the degree of deviation through the relation

\begin{equation}\label{erro}
D(\%) = \frac{\vert\dot\rho_{\star}/\langle\varepsilon\rangle - \dot\rho_{\rm SFR}\vert}{\dot\rho_{\star}/\langle\varepsilon\rangle}\times 100\%,
\end{equation}

\noindent taking the distribution of deviations in relation to the total number of points within three classes: the first class encompassing deviations less than 1\%; the second class comprising deviations between 1\% and 5\%; the last class considering deviations between 5\% and 10\%.

From Figures \ref{fig4acde} and \ref{fig5abcd} (and also Table \ref{err}), it is possible to verify the good mapping that the SFR, through the use of both isothermal and non-isothermal PDFs, has made of the CSFR since the time when the first star formed in the universe ($z\sim 20$) up to the present.

\begin{table}[h!]
\begin{center}
\caption{Mach Numbers for Each of the Nine Models Analyzed}
\begin{tabular}{cccccccccccccccccccc}
\hline 
\hline 
& & & & \multicolumn{4}{c}{$s_{\rm crit} = 2$} & & &  \multicolumn{3}{c}{$s_{\rm crit} = 3$} & & & & \multicolumn{3}{c}{$s_{\rm crit} = 4$} \\
\hline
$\Gamma$ & & & & $z=0$ & & $z=3.5$ & & & & $z=0$ & & $z=3.5$ & & & & $z=0$ & & $z=3.5$ & \\
1.0 & & & & 6.8 & & 5.8 & & & &  10.9 & & 9.3 & & & & 20.8 & & 16.6 \\
1.1 & & & & 7.6 & & 6.4 & & & &  13.8 & & 11.3 & & & & $...$ & & $...$ \\
1.2 & & & & 8.7 & & 7.0 & & & &  21.8 & & 15.2 & & & & $...$ & & $...$ \\
1.3 & & & & 10.2 & & 7.9 & & & & $...$ & & $...$ & & & & $...$ & & $...$ \\
1.4 & & & & 12.8 & & 9.3 & & & & $...$ & & $...$ & & & & $...$ & & $...$ \\
\hline
\end{tabular}
\label{resumo}
\end{center}
{\bf Note.} The values of $\mathcal{M}$ are identified in two distinct instants of time. The redshift $z = 3.5$ (the universe is about $1.8\,{\rm Gyr}$ old for the cosmological parameters used to characterize the CSFR) corresponds to the instant of time when the CSFR reaches the peak while $z = 0$ ($\sim 13.7\,{\rm Gyr}$ for the age of the universe) represents the local universe.
\end{table}

\begin{table}[!ht]
\begin{center}
\caption{Distribution of the Deviations of the Equality Established by Equation (\ref{igual2}) When We Divide the Interval in Redshift ($0-20$) into 12,000 Linearly Spaced Points}
\centering
\begin{tabular}{ccccccccc}
\hline
\hline 
$s_{\rm crit}$ & & $\Gamma$ & & $D < 1\%$ & & $1\%\leq D < 5\%$ & & $5\% \leq D < 10\%$ \\ 
\hline 
2 & & 1.0 & & 0.980 & & 0.020 & & $8.33\times 10^{-5}$ \\ 
2 & & 1.1 & & 0.982 & & 0.018 & & $...$  \\
2 & & 1.2 & & 0.983 & & 0.017 & & $...$  \\
2 & & 1.3 & & 0.998 & & 0.002 & & $...$  \\
2 & & 1.4 & & 0.999 & & 0.001 & & $...$  \\
3 & & 1.0 & & 0.994 & & 0.006 & & $...$  \\
3 & & 1.1 & & 0.999 & & 0.001 & & $...$  \\
3 & & 1.2 & & 0.999 & & 0.001 & & $...$  \\
4 & & 1.0 & & 0.999 & & $...$ & & $8.33\times 10^{-5}$ \\ 
\hline
\end{tabular} 
\label{err}
\end{center}
{\bf Note.} For all nine models, the deviations presented can be considered very small---less than 1\% for more than 98\% of the points considered in the analysis.
\end{table}

Nevertheless, this mapping cannot be performed for any values of $s_{\rm crit}$ and $\Gamma$, as is clear from the absence of specific models in Tables \ref{resumo} and \ref{err}. For example, if we take the $s_{\rm crit}=2$ model with $\Gamma = 5/3$, it will be possible to keep the Equation (\ref{igual2}) valid from $z=20$ to $\sim 6$. From $z <6$, the mapping of the CSFR by the SFR breaks and the equality represented by Equation (\ref{igual2}) is no longer valid. In particular, the integral in the Equation (\ref{igual2}) does not provide sufficient ``power," through the Mach number, to cover the variation of the $\rho_{\star}$ and $\langle\varepsilon\rangle$ parameters that are on the left side of the equality. Thus, a full map over the entire range in redshift can not be obtained.

Specifically for the non-isothermal PDF, the maximum and minimum values for the Mach number, which can be applied to provide the solution of the integral (\ref{sfrhopkins}), are limited by the condition $\omega (s)\geq 0$. Thus, the models presented are those that effectively allow a complete mapping of the CSFR through the SFR within the entire range in redshift. All models that fail to make the complete CSFR map have similar characteristics. That is, they can properly map the CSFR from $z=20$ to intermediate redshifts ($\sim 7-4$), but fail on the $z\lesssim 6$ scale. As our main objective in this paper is to analyze the complete mapping between the CSFR and the SFR, we do not link $s_{\rm crit}$ through the $s_{\rm crit} \sim {\rm ln}(\alpha_{\rm vir}\mathcal{M}^{2})$ relation. The influence of the virial parameter on the results of this unified model will be explored in another publication.

\subsection{Relationship between Mach Number and Star Formation Efficiency}\label{sfrmachcsfr}

\begin{figure}[ht!]
\epsscale{1.15}
\plottwo{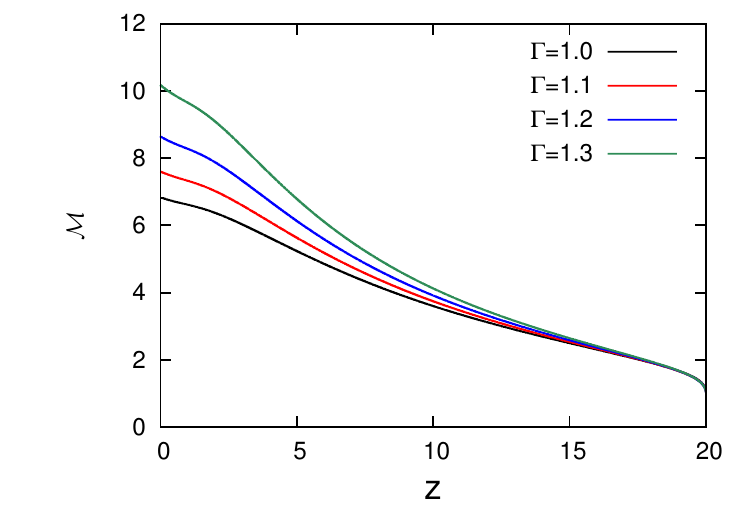}{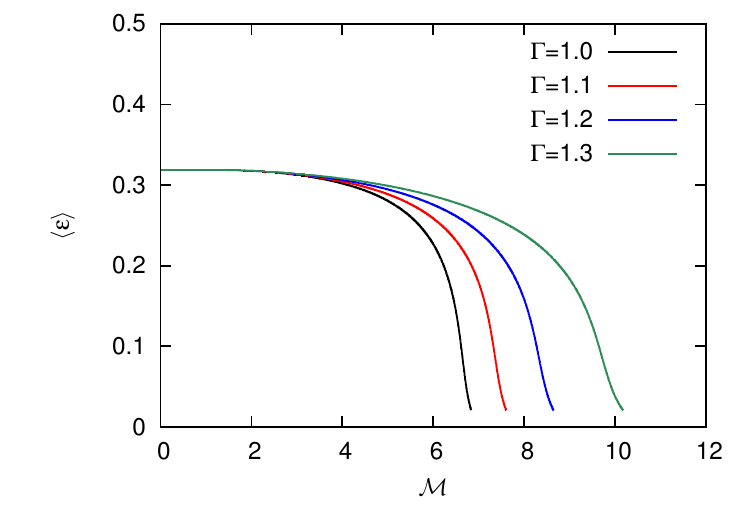}
\plottwo{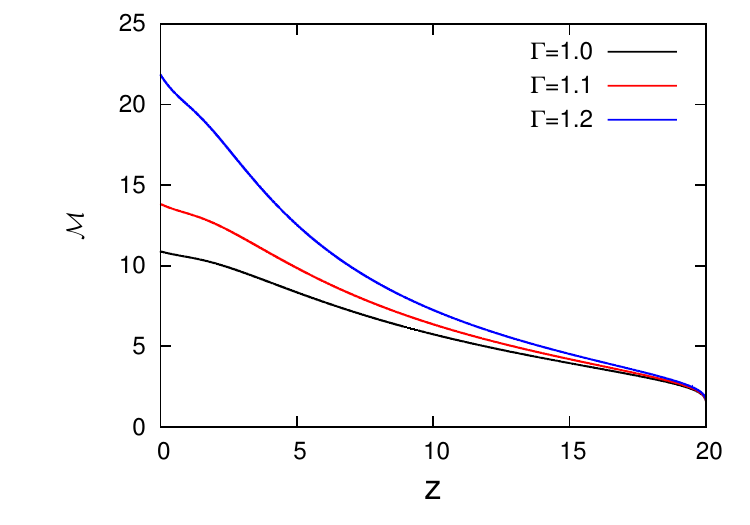}{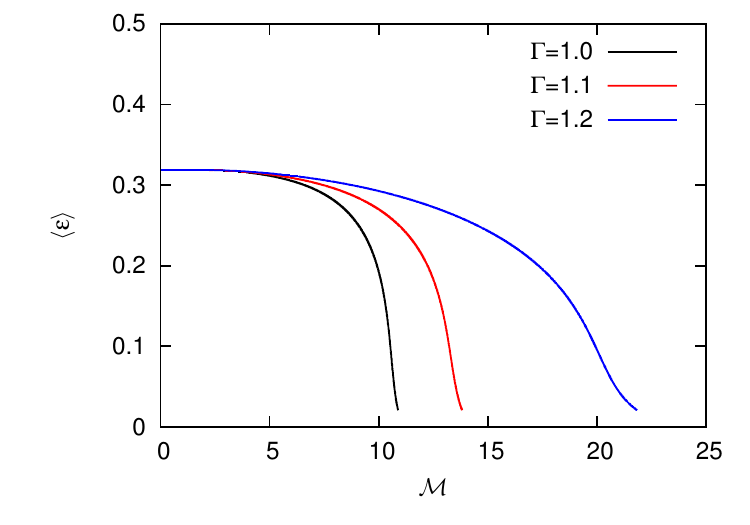}
\plottwo{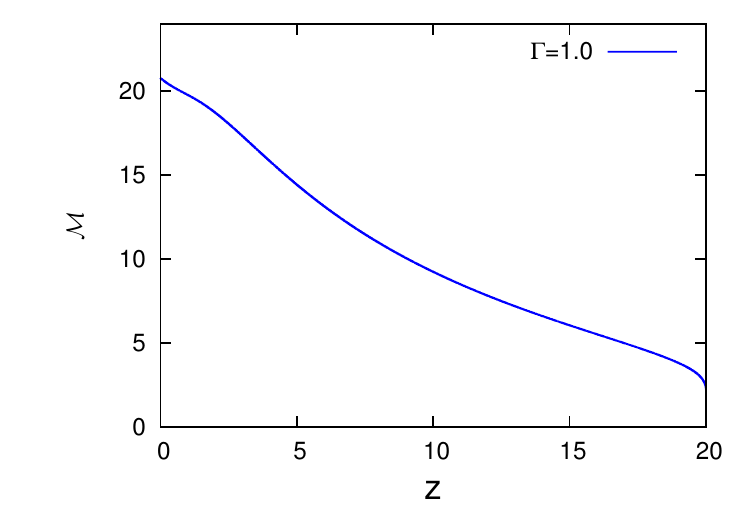}{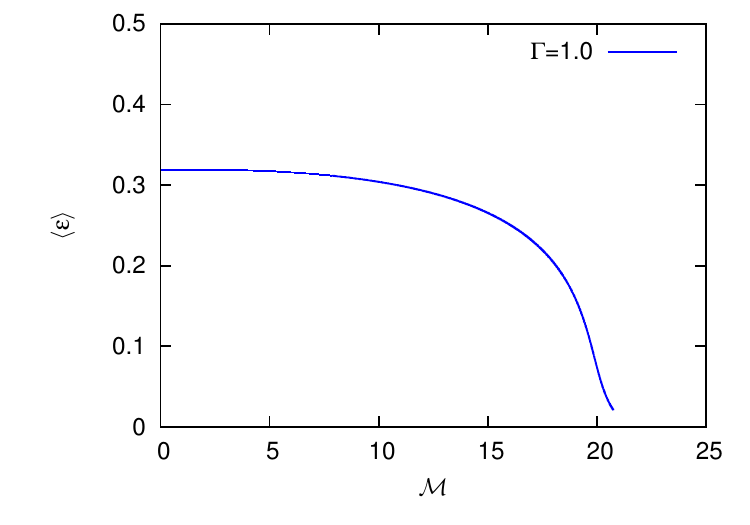}
\caption{The curves show the evolution of the Mach number with the redshift from the map generated by the SFR for the CSFR, and also show how the cosmological star formation efficiency, $\langle\varepsilon\rangle$, is linked to $\mathcal{M}$. The top panels show some models identified by the values of $\Gamma$ with $s_{\rm crit} = 2$. Intermediate panels show the results for $s_{\rm crit} = 3$. The bottom panels show the $s_{\rm crit} = 4$ model with $\Gamma = 1$.\label{fig676a7a}}
\end{figure}

In the Figure \ref{fig676a7a}, we present the evolution of $\mathcal{M}$ with the redsfhift and also how the star formation efficiency $\langle\varepsilon\rangle$ varies with the Mach number. These results derive directly from the mapping of the CSFR by the SFR. The upper panels show some models identified by the values of $\Gamma$, considering $s_{\rm crit} = 2$, the bottom panels show the results for the model with $s_{\rm crit} = 4$ and $\Gamma = 1$, while the middle panels show some models with $s_{\rm crit} = 3$.

It is important to emphasize, once again, the complementary character that exists between the two rates of star formation. The identification that the SFR, through the PDF commonly used to study the giant molecular clouds in our Galaxy, can effectively mimic the behavior of the CSFR, from the time the first star formed in the universe to the present, allows us to infer the role of Mach number, and therefore of the turbulence, in the formation of the large-scale structures of the universe.

Certainly, our formalism lacks the ability to provide rich details, as large computational simulations do. However, our results represent average values weighted by the mass of the dark matter halos that host the baryonic matter that is the basic material for star formation. By analyzing the panels on the left side of the Figure \ref{fig676a7a} ($\mathcal{M}$ versus $z$), we can verify that when the first set of halos forms at $z = 20$, generating the potential wells for the fall of the baryonic matter, the star formation begins with a low Mach number.

In the hierarchical structure formation scenario, the first halos have masses $\sim 10^{6}M_{\odot}$. As the redshift decreases, more and more massive halos are able to decouple from the Hubble flow, collapse, and virialize, generating conditions to capture more and more baryons from the surrounding environment (the universe itself). Thus, the Mach number increases with the growth of the CSFR. In the case $s_{\rm crit} = 2$, there is no great influence of the polytropic index ($\Gamma$) on the results up to $z\sim 12$. For the case $s_{\rm crit} = 3$, we can verify that a $z$ value up to $\sim 15$ does not observe great influence of the polytropic index, and the models differ little. However, as the universe evolves, the $\Gamma$ parameter becomes more important to the value of $\mathcal{M}$.

The increase of $\mathcal{M}$ with the value of $\Gamma$ is consistent with the formalism presented in section \ref{subsec:sfr}, and synthesized through Equations (\ref{iso1})--(\ref{iso3}), as well as from the analysis of several authors with respect to the relation $\sigma_{\rm s}$ versus $\mathcal{M}$ (see, e.g., \citeauthor{federrath2015} \citeyear{federrath2015}; \citeauthor{nolan2015} \citeyear{nolan2015}). Another aspect associated with the $\mathcal{M}$ versus $z$ relation is that our results for both $z\sim 1-3$ and $z = 0$ typically correspond to the average values obtained by \cite{salim2015}. In that paper, the authors present predictions of the Mach number for extragalactic sources. Our results agree with the estimates of these authors for the disk galaxies (see Table 3 of these authors). A similar result is obtained from the comparison of our model with that studied by \cite{renaud2012}. The authors find that $\mathcal{M}=10$ for disc galaxies at high redshifts (see, in particular, Figure 4 of these authors), a result that is consistent with the results derived by \cite{salim2015} and those obtained here. In addition, our results with $s_{\rm crit} = 3$ ($\Gamma=1.2$) and $s_{\rm crit} = 4$  marginally return the estimates for the Mach number from \cite{renaud2012} to high-$z$ mergers.

Looking at the right-hand panels of Figure \ref{fig676a7a}, we observe the efficiency behavior associated with the star formation process, generated by the CSFR, versus $\mathcal{M}$ provided by the SFR. All models show similar characteristics, with a high star formation efficiency, $\langle\varepsilon\rangle$, up to a certain $\mathcal{M}_{\rm crit}$. From this critical value, the star formation efficiency rapidly decreases. This shows the dual role played by turbulence as proposed by \cite{klessen2010}. The same authors argue that the formation of the first stars of the universe were subject to the same dynamic processes of the local star-forming regions. This is exactly the result described from the mapping CSFR--SFR.

Note that, for $\mathcal{M} <\mathcal{M}_{\rm crit}$, the results do not depend significantly on the polytropic index, while for $\mathcal{M}> \mathcal{M}_{\rm crit}$, in addition to a rapid decrease of $\langle\varepsilon\rangle$, there exists a differentiation between the different $\Gamma$s in the results $\langle\varepsilon\rangle$ versus $\mathcal{M}$. The higher the value of $\Gamma$, the greater the value of $\mathcal{M}$ associated with a given efficiency will be. In particular, see the dependency that also exists with $s_{\rm crit}$. That is, $\mathcal{M}_{\rm crit} \sim 4$ for $s_{\rm crit}=2$, while for $s_{\rm crit}=3$ we have $\mathcal{M}_{\rm crit} \sim 6$, and for $s_{\rm crit}=4$ we find $\mathcal{M}_{\rm crit} \sim 8$.

\cite{klessen2000} showed that the star formation efficiency decreases systematically as either the driving scale of the turbulence is decreased or the turbulent Mach number is increased. In particular, our unified model shows this behavior when $\mathcal{M}$ exceeds $\mathcal{M}_{\rm crit}$. It is worth stressing that \cite{federrath2015} present an interesting analysis of the structures formed from non-isothermal polytropic turbulence. The authors find, as a result of their simulations, that $\Gamma < 1$ leads to a more fragmented density field with filaments with high density contrasts, while $\Gamma > 1$ softens the density contrasts of small scales. Observing Figure 3 of \cite{federrath2015}, especially the intermediate panel showing the volume-weighted Mach number versus time ($t/T$, where $T$ is the turbulent crossing time), it is possible to see that higher values of $\Gamma$ allow to reach higher values for $\mathcal{M}$ for the same time $t / T$. Note that our results presented in Figure \ref{fig676a7a} for $\mathcal{M}$ versus $z$, considering different $\Gamma$ values, show similarity with these results presented by \cite{federrath2015}. In particular, the redshift is a parameter directly associated with $t$. Thus, $ z \rightarrow 20 $ represents the case $t / T \rightarrow 0$ of these authors, where $\mathcal{M}$ is practically insensitive to the $\Gamma$ value. Below a given redshift, larger values for the $\Gamma$ parameter produce higher values for $\mathcal{M}$, a result that is analogous to that of these authors for $t / T> 0.5$ (corresponding, for example, to $z <12$ for models with $s_{\rm crit} = 2$ or $z <15$ for models with $s_{\rm crit} = 3$).

Another interesting comparison of relation $\langle\varepsilon\rangle$ versus $z$ (Figure \ref{fig:e_csfr}), which in our case allows SFR to map the relation $\mathcal{M}$ versus $z$, can be made with the recent work of \cite{scoville2017}. These authors, using ALMA observations from the long wavelength dust continuum, estimate ISM masses for 708 galaxies within the range $\sim 0.3-4.5$ in redshift. In that work, they show the evolution of the stellar formation efficiency (SFE in the nomenclature of those authors) within the range $0-3.5$ and through the relative ratio ${\rm SFE}(z)/{\rm SFE}(z = 0)$. We observe that our ratio $\langle\varepsilon(z)\rangle / \langle\varepsilon(z=0)\rangle$  is greater than that by approximately a factor $\sim 1-2.5$ within the same range $0-3.5$. \cite{scoville2017} conclude that the increase in the star formation within the analyzed redshift range is due to both the increase in mass of the ISM and the increase in the conversion of gas to stars. This result is identically obtained by PM in their model for the CSFR.

The discussions presented in this section reinforce our analysis of the complementarity between the CSFR and the SFR, observed through the ``SFR's mimicry." In particular, these results allow us to conclude that the relations $\mathcal{M}$ versus $z$ and $\langle\varepsilon\rangle$ versus $\mathcal{M}$ derived from our analysis are perfectly consistent with the unified model here presented, in addition to representing well the physical processes that have been discussed by different authors in recent works on the SFR.

\subsection{CSFR Providing the Larson's First Law for the SFR}\label{csfrlarsonlaw}

In the previous section, we have seen the contribution that the formalism used for the SFR can contribute to the CSFR, nominally, Mach number relations that provide both the redshift and the cosmic efficiency of star formation. These relationships can not be directly obtained from the formalism used by PM without the help of the SFR. In contrast, in this section, we show a contribution, which can be provided directly by the CSFR to the SFR, that is a way of providing Larson's first law. Rewriting Equation (\ref{st}) in the form

\begin{equation}\label{st2}
n(M,z) dM = f(\sigma,z)\frac{\rho_{\rm B}}{M^{2}}\frac{d\left[{\rm ln}\sigma(M,z)\right]}{d{\rm ln}M}dM,
\end{equation}

\noindent where the variables of Equation (\ref{st2}) were defined in section \ref{sec:sfr}, it enables us to estimate the average mass of the halos formed as a function of the redshift using the scenario proposed by PM for the CSFR. This can be done through

\begin{equation}\label{masshalo}
\langle M_{\rm H}(z)\rangle = \frac{\rho_{\rm B}(z)}{\int_{M_{\rm min}}^{\infty} n(M,z) d{\rm ln}M}.
\end{equation}

In the theory of cosmological perturbations, fluctuations in the dark matter begin to grow after equipartition\footnote{In fact, perturbations in the dark matter can grow even during the time when radiation dominates. However, in this case, the density contrast is $\delta_{\rm dm}\propto {\rm ln}(t)$. After equipartition, the growth of the dark matter density contrast becomes $\delta_{\rm dm} \propto t^{2/3}$. On the other hand, the baryonic density contrast increases only after recombination ($z\sim 1100$), when baryons decouple from the radiation.} (the instant of time when the densities of matter and radiation become equal). As they evolve, the perturbations in the dark matter expand with the Hubble flow in an increasingly slower way. Upon reaching density contrast $\delta_{\rm c} \sim 1.69$, the perturbations detach from the expansion of the universe and collapse. Because dark matter is not dissipative, the collapse stops when the density contrast reaches a value of $\sim 200$. This value represents the condition called virialization of the halos. Thus, we can estimate the average virial radius, associated with $\langle M_{\rm H}(z)\rangle$, through

\begin{equation}\label{virial}
\langle R_{\rm V}(z)\rangle = \left(\frac{3\,\langle M_{\rm H}\rangle}{800\pi\rho_{\rm B}}\right)^{1/3}.
\end{equation}

As the baryonic matter is dissipative, it will tend to cluster more in the interior of the halos. Our {\it ansatz} in this case is to consider that all the gas ($\rho_{\rm g}$) is distributed within radius ($\langle R_{\rm V}(z)\rangle$). However, the part of the gas that will produce stars ($\rho_{\rm mol}$) will reside in the innermost part of the halos generating the density of stars $\rho_{\star}$ within an effective radius $R_{\star}$, which can be estimated by

\begin{equation}\label{raiobarions}
\langle R_{\star}(z)\rangle = \frac{\rho_{\star}}{\rho_{\rm g}} \langle R_{\rm V}(z)\rangle,
\end{equation}

Equation (\ref{raiobarions}) should be seen as an initial proposal (toy model) in order to verify the possibility of Larson's law emerging from this formulation. In addition, very probably, there are a large number of fusions of low-mass halos generating higher-mass halos. From this rich environment could emerge a scale relation between $\langle R_{\rm V}\rangle$ and $\langle R_{\star}\rangle$, similar to that proposed by the equation above.

All of these phenomena are likely to contribute to the gas on the large scale; in this case, the large scale corresponds to $\langle R_{\rm V}\rangle$, transferring kinetic energy to the star-forming gas ($\rho_{\rm mol}$) that lies in the innermost part of the halos. As a result the star-forming gas will produce stars within an effective radius $\langle R_{\star}\rangle$, whose density of stars formed will be $\rho_{\star}$ (converting from $\rho_{\rm mol}$ to $\rho_{\star}$ on a characteristic time scale $\tau_{\rm s}$).

The key point of the present analysis is that the mapping described by Equation (\ref{igual2}) must be valid in both directions. That is, if the SFR can, through the Mach number, appropriately map the CSFR by allowing parameters such as efficiency of the cosmological star formation (which is related to the redshift in the cosmological context) can be associated with the Mach number, then it must also be possible that the CSFR can map the SFR through the characteristic scale $\langle R_{\star}\rangle$ in which the formation of stars regulated by the turbulence occurs.

Following, for example, \cite{hennebelle2008,hennebelle2009}, we wrote for the gas velocity dispersion

\begin{equation}\label{vrms}
\langle V_{\rm rms}^{2}\rangle = \mathcal{M}^{2}c_{\rm s}^{2},
\end{equation}

\noindent where $c_{\rm s}$ represents the thermal sound speed. Considering a polytropic equation of state and that the gas behaves as a perfect gas, we have

\begin{equation}
P = \kappa \rho_{\rm mol}^{\Gamma} = \frac{k_{\rm B}}{\mu m_{\rm H}}\rho_{\rm mol} T(\rho_{\rm mol}),
\end{equation}

\noindent from which we obtain

\begin{equation}
\kappa = \frac{k_{\rm B}}{\mu m_{\rm H}}\rho_{\rm mol}^{1-\Gamma}T(\rho_{\rm mol}).
\end{equation}

Thus, the temperature depends on the density via

\begin{equation}
T(\rho_{\rm mol}) = T_{0} \left(\frac{\rho_{0}}{\rho_{\rm mol}}\right)^{1-\Gamma},
\end{equation}

\noindent and the thermal sound speed can be written as

\begin{equation}
c_{\rm s} = \sqrt{\frac{\partial P}{\partial \rho_{\rm mol}}} = \left({\Gamma \frac{k_{\rm B}}{\mu m_{\rm H}}\rho_{0}^{1-\Gamma}T_{0}}\right)^{1/2}\,\rho_{\rm mol}^{(\Gamma-1)/2},
\end{equation}

\noindent where $k_{\rm B}$ is the constant of Boltzmann, $m_{\rm H}$ is the mass of the hydrogen atom, $\mu\sim 0.5$ is the average molecular weight of the gas, and $\rho_{0}$ and $T_{0}$ correspond to the average values for the gas density and temperature, respectively.

\begin{figure}[ht!]
\epsscale{1.15}
\plottwo{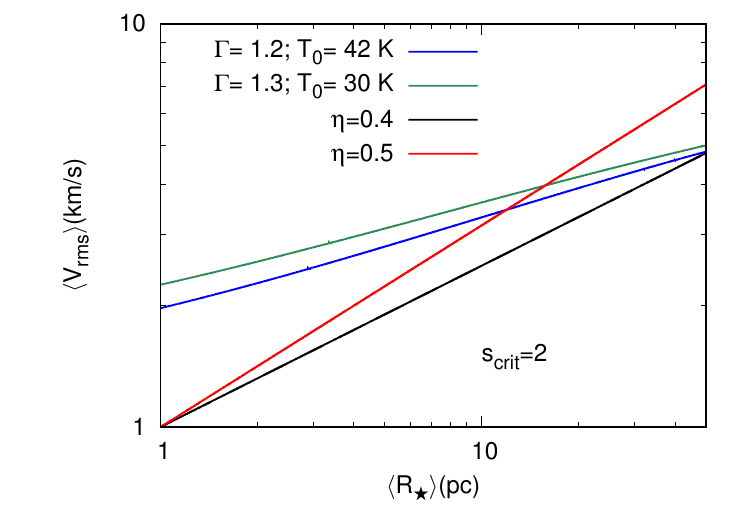}{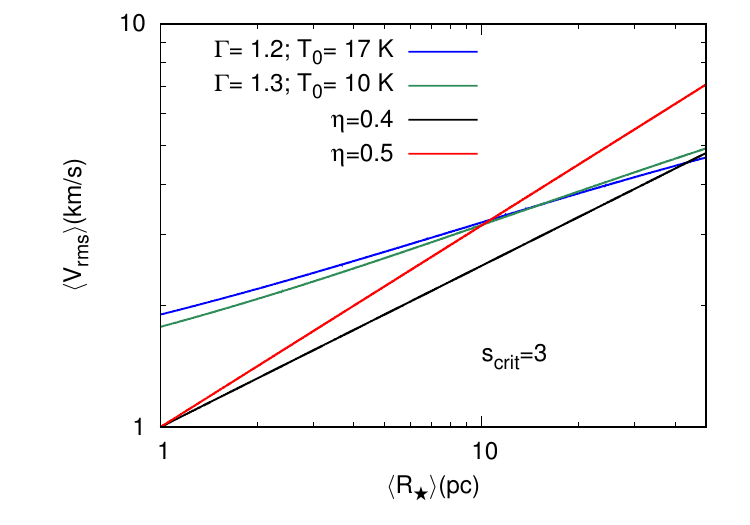}
\includegraphics[scale=1.15]{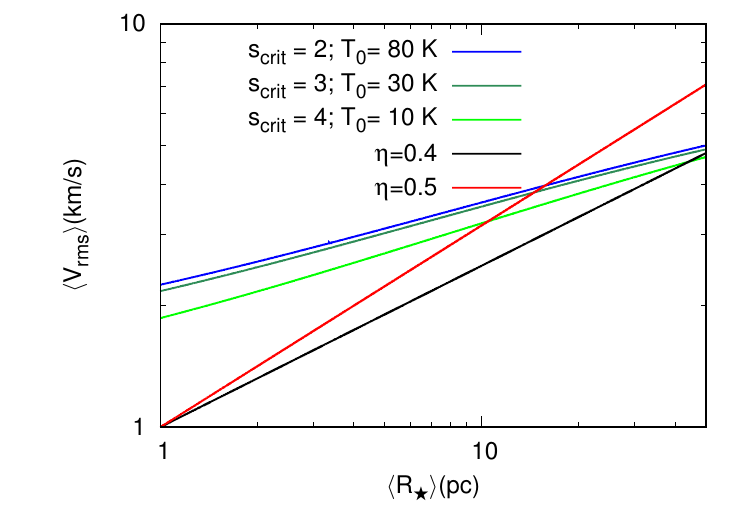}
\caption{The panels show the solutions for the relationship $\langle V_{\rm rms}\rangle$ versus $\langle R_{\star}\rangle$, considering different values for both the polytropic index and $s_{\rm crit}$. The temperature $T_{0}$ is obtained from the value that produces the best fit for the Larson's law as defined by Equation (\ref{larsonlaw}), for the limit values $\eta = 0.4-0.5$. The bottom single panel correspond to the models with $\Gamma = 1$.\label{fig898a9a8b9b}}
\end{figure}

Because the gas falls into the gravitational potential wells of the halos, it will tend to distribute within $\langle R_{\rm V}\rangle$, generating an average density $\rho_{0}$. The estimate for the value of this parameter can be obtained from $s={\rm ln}(\rho_{\rm g}/\rho_{0})$. Assuming that, within $\langle R_{\rm V} \rangle$, the gas has a typical density contrast of the order of $s_{\rm crit}$ ($\langle s\rangle \simeq s_{\rm crit}$), as  a characteristic value, then it is possible to express $\rho_{0}$ as a function of $\rho_{\rm g}$ and as a function of $\rho_{\rm mol}$. Defining the value of the characteristic temperature, $T_{0}$, we can calculate the thermal sound speed as a function of redshift. As we have the solution $\mathcal{M}$ versus $z$, for each specific value of $\Gamma$, obtained from the mapping of the CSFR by the SFR, it is thus possible to calculate $\langle V_{\rm rms}\rangle$ in Equation (\ref{vrms}).

Once $\rho_{\rm mol}$ is converted to $\rho_{\star}$ on the scale $\langle R_{\star}\rangle$, then we can construct the solution $\langle V_{\rm rms}\rangle$ versus $\langle R_{\star}\rangle$. If the inverse mapping can be done, then it will be possible to compare $\langle V_{\rm rms}\rangle$ versus $\langle R_{\star}\rangle$ with the Larson's first law represented by Equation (\ref{larsonlaw}). The last step is to vary the parameter $T_{0}$ in order to obtain the best possible adjustment of the $\langle V_{\rm rms}\rangle - \langle R_{\star}\rangle$ to the limits given by Larson's law within the range $\sim 1-50\,{\rm pc}$. The result of this analysis is shown in Figure \ref{fig898a9a8b9b}.

The panels at the top of Figure \ref{fig898a9a8b9b} show the models identified by their $\Gamma$ values, for $s_{\rm crit} = 2$ (left) and $s_{\rm crit} = 3$ (right), while the lower panel shows the results for $\Gamma = 1$ with $s_{\rm crit} = 2$, $3$, and $4$. The results are dependent on both the polytropic index and the $s_{\rm crit}$ values. For $\Gamma = 1.2-1.3$, the change from $s_{\rm crit} = 2$ to $3$ allows to reduce the value of $T_{0}$ by a factor $\sim 3$. None of the models studied fit the Larson's law well for $R_{\star}<10\,{\rm pc}$, although at the scales closest to $10\,{\rm pc}$, the model curves tend to approximate the $R^{0.5}$ law. In the range of $10-50\,{\rm pc}$, all models remain within the bounds $\eta\simeq 0.4-0.5$ and are approaching the curve $R^{0.4}$ on the larger scales. These results show that, in principle, it would be possible to use for CSFR to obtain the Larson's law on $\sim 10-50\,{\rm pc}$ scales. All models have temperatures $T_{0}$ within the range $\sim 10-80\,{\rm K}$.

Recently, \cite{tang2017a} presented measurements of kinetic temperature for six different regions of star formation in the Large Magellanic Cloud (LMC). Because it is a nearby galaxy in a low-metalicity environment, it is likely that the star-forming regions studied by these authors may be more representative of the model we present in this section. Using non-local thermodynamic equilibrium (NLTE) models, \cite{tang2017a} obtain kinetic temperatures within the interval $\sim 25-80\,{\rm K}$ with 30 Dor the source presenting the highest sample temperature. Similar results can be observed in \cite{tang2017b}, who obtain kinetic temperatures $\sim 30-61\,{\rm K}$, for massive star forming molecular clumps, from para -- ${\rm H_{2}CO}$ ($3_{21} -2_{20} / 3_{03} -2_{02}$) lines ratio. These results are compatible with the results achieved in our work.

It is worth mentioning that the hierarchical structure formation scenario predicts the existence at $z = 0$ of a large number of low-mass halos that are not directly observed. This can be explained in two different, non-exclusive ways. The first considers the observational bias associated with the limit of detection of objects with low luminosity in a given sample. The second possibility is associated with the fusion of low-mass halos, or their incorporation by much more massive halos. In the second case, massive halos could be composed of a number of low-mass mini-halos. From the way we map the SFR to get Larson's law, the hypothesis that mini-halos can be embedded by halos of greater mass is implicit. In principle, these mini-halos would contain a certain number of stars in a similar way to the one that is verified, mainly, in the globular clusters (GCs).

In a recent study, \cite{sollima2016} estimate the fraction and distribution of dark matter in the innermost regions of two GCs of the Milky Way, namely NGC 6218 (M12) and NGC 288. The authors estimate that there is a large mass fraction in these clusters that is compatible with concentrated non-luminous matter. More recently, \cite{penarrubia2017} have shown that encounters in the central regions of GCs embedded in dark matter halos necessarily lead to the formation of an equilibrium configuration that extends far beyond the stellar radius of the GCs. In particular, with $M_{\rm DM}\sim 10^ {6}M_{\odot}$, the authors find that the distribution of stars could reach hundreds of parsecs while keeping their equilibrium configurations. In addition, the presence of dark matter may lead to an increase in the line-of-sight velocity dispersion of these systems.

\section{Summary and Conclusions} \label{sec:conc}

We present a unified model that allows us to describe both the cosmological star formation represented by the CSFR and the local star formation represented by the SFR. Due to its healthy characteristics, we use the formulation proposed by \cite{pereira2010} to describe the CSFR, while the SFR is described by the formulation discussed in \citeauthor{hopkins2013a} (\citeyear{hopkins2013a}, \citeyear{hopkins2013b}) and \cite{federrath2015}. The central point of our analysis is synthesized in Equation (\ref{igual2}), which in turn allows, as an {\it anstaz}, that the variations of $\dot\rho_{\star}/\langle\varepsilon\rangle$ with the redshift can be mapped by the Hopkins (general case) or isothermal PDFs through the Mach number ($\mathcal{M}$). Complete mappings from redshift $\sim 20$ to the present can only be obtained for certain combinations of $s_{\rm crit}$ and $\Gamma$ (keeping in mind that the connection between Hopkins' PDF and $\Gamma\neq 1$ was established by \citeauthor{federrath2015} \citeyear{federrath2015}). Looking at the results presented through Figures \ref{fig4acde} and \ref{fig5abcd} in addition to Table \ref{err}, we can conclude that the PDFs ordinarily used for studying the formation of stars in our Galaxy and the near universe can effectively mimic the CSFR, which in turn is constructed from the hierarchical structure formation scenario. Our main conclusions are:

(i) Star formation begins at high redshifts ($z\sim 20$), with gas presenting low Mach numbers (subsonic scale $\mathcal{M} \sim 0.5$). The first stars of the universe are formed in halos of dark matter with typical masses $\sim 10^{6}-10^{7}M_{\odot}$.

(ii) As the number of halos of higher mass increases, with the reduction of redshift, more baryonic matter falls into the wells of gravitational potential generated by these structures. The density of both the gas and the stars increases, causing the degree of gas turbulence parameterized by $\mathcal{M}$ to increase as well. For $\mathcal{M} \lesssim 3-4$, the results are little influenced by the value of the polytropic index ($\Gamma$).

(iii) Within the \cite{pereira2010} formulation for the CSFR, $\dot\rho_{\star}$ reaches its maximum value close to redshift $\sim 3.5$ and the SFE ($\langle\varepsilon\rangle = \rho_{\rm mol}/ \rho_{\rm g}$) varies little within the $z\sim 3.5-20$, being close to $\langle\varepsilon\rangle\sim 0.3$ in that interval. At $z = 3.5$, the Mach number reaches a value for $s_{\rm crit} = 2$ given by the relation $\mathcal{M}_{\rm crit} \sim 5.8\,\Gamma^{1.2}$ ($\Gamma\leq 1.3)$, while $\mathcal{M}_{\rm crit} \sim 5.8\,\Gamma^{1.4}$ best describes the Mach number for $\Gamma = 1.4$. For $s_{\rm crit} = 3$, we find $\mathcal{M}_{\rm crit} \sim 9.3\,\Gamma^{2.65}$. The $s_{\rm crit}=4$ model can map the two star formation rates only to $\Gamma = 1$; in this case, $\mathcal{M}_{\rm crit} \sim 16.6$. For $\mathcal{M} < \mathcal{M}_{\rm crit}$, the star formation efficiency is high and almost constant. Above $\mathcal{M}_{\rm crit}$, the efficiency drops rapidly as $\mathcal{M}$ grows.

(iv) Because the CSFR provides $\langle\varepsilon\rangle$ versus $z$ while the SFR provides the Mach number, it is possible to construct the relations  $\langle\varepsilon\rangle$ versus $\mathcal{M}$ and $\mathcal{M}$ versus $z$. In particular, the identified behavior of the relation $\mathcal{M}$ versus $z$, as a function of different polytropic indices, is similar to that observed from \cite{federrath2015} simulations and related to the volume-weighted Mach number versus time (where time is parameterized as $t / T$, with $T$ the turbulent crossing time).

(v) At $z = 0$, the typical values of $\mathcal{M}$ lie between $\sim 7-13$ for $s_{\rm crit} = 2$, $\mathcal{M}\sim 11-22$ for $s_{\rm crit} = 3$ and $\sim 21$ for $s_{\rm crit} = 4$. Considering $\mathcal{M} = 10$ as the typical value for the Milky Way (see \citeauthor{federrath2015} \citeyear{federrath2015} and references therein), our results are close to this value, at $z=0$, for most of the nine models analyzed in this work. Another point is that our results for $\mathcal{M}$ versus $z$ for both $z\sim 1-3$ and $z = 0$ typically correspond to the mean values obtained by \cite{salim2015} for disk galaxies (similar result for the sample of disk galaxies analyzed by \citeauthor{renaud2012} \citeyear{renaud2012}). In addition, our results with $s_{\rm crit} = 3$ ($\Gamma=1.2$) and $s_{\rm crit} = 4$  marginally return the estimates for the Mach number from \cite{renaud2012} to high-$z$ mergers.

(vi) The turbulence shows a dual character, inducing the star formation with high values of $\langle\varepsilon\rangle$, until reaching $\mathcal{M}_{\rm crit}$. For $\mathcal{M} > \mathcal{M}_{\rm crit}$, a strong decrease in the SFE occurs. Thus, turbulence is a regulator of the star formation, playing the dual role proposed by \cite{klessen2010}.

(vii) The ratio $\langle\varepsilon(z)\rangle / \langle\varepsilon(z=0)\rangle$ provided by PM-CSFR model is in good agreement with that obtained by \cite{scoville2017}, within the redshift range $0-3.5$.

(viii) \cite{pereira2010} in their work argue that $\tau_{\rm s}\sim 2\,{\rm Gyr}$, with a Salpeter exponent, provides good agreement with the observational data of the CSFR. With this value for $\tau_{\rm s}$, we obtain $\langle\varepsilon\rangle = 0.021$ at $z=0$, which is comparable with $\varepsilon_{\rm ff} \sim 0.01$ and $\tau_{\rm dep}\sim 1-2.2\,{\rm Gyr}$, as inferred by several authors for star-forming regions in our Galaxy (see, e.g., \citeauthor{krumholz2005} \citeyear{krumholz2005}).

(ix) Using the CSFR as a map for the SFR, it is possible to obtain a relation for the velocity dispersion of the gas that will be directly involved with the star formation within the dark matter halos. In this case, following the works of \cite{hennebelle2008,hennebelle2009}, we show that Larson's first law can be consistently obtained. The inferred temperatures in our model are within the range $\sim 10-80\,{\rm K}$, which are values similar to those inferred by authors such as \citeauthor{tang2017a} (\citeyear{tang2017a}, \citeyear{tang2017b}) for molecular clouds of our Galaxy and for the LMC. We restrict our analysis to the $\sim 1-50\,{\rm pc}$ range. Although the fit for Larson's law is not good in the $\sim 1-10\,{\rm pc}$ range, our model shows consistency with Equation (\ref{larsonlaw}), particularly for $\langle R_{\star}\rangle \sim 10-50\,{\rm pc}$.

(x) The formulation that allows to obtain the Larson's law implicitly adds the hypothesis that the halos of greater mass are composed of a number of halos with much smaller masses. Thus, the cosmological star formation would be processed, in part, in structures similar to globular clusters. The presence of non-baryonic dark matter in globular clusters has recently been discussed by \cite{sollima2016} and \cite{penarrubia2017}. Our work shows consistency with the results and analyses of these authors.

Our study demonstrates that there is strong complementarity between the formulations used to derive the CSFR and the SFR, so that it is possible to think of a unified model that adequately describes both cosmological and Galactic star formation. Although our model is semi-analytical, and therefore cannot provide rich details like those obtained from computational simulations, it can provide several interesting clues about the role of turbulence as a regulator of star formation, as well as the existence  of an $\mathcal{M}_{\rm crit}$ from which the efficiency of star formation rapidly decreases. In addition, our model identifies the role of Larson's first law as a result of the very formation of large-scale structures of the universe, which in turn would allow the formation of galactic systems including our Galaxy.

\acknowledgments

We thank the anonymous referee for useful comments, which improved this work. We would also like to thank the Brazilian agency FAPESP for support under thematic project 2014/11156-4. C.G would like to thank CAPES for a graduate research fellowship. O.D.M. thanks CNPq for partial financial support (grant 303350/2015-6).


\end{document}